\documentclass[acmsmall]{acmart}

\setcopyright{cc}
\setcctype{by}
\acmDOI{10.1145/3747515}
\acmYear{2025}
\acmJournal{PACMPL}
\acmVolume{9}
\acmNumber{ICFP}
\acmArticle{246}
\acmMonth{8}
\received{2025-02-27}
\received[accepted]{2025-06-27}

\usepackage{booktabs}   
\usepackage{subcaption} 
\usepackage{anyfontsize}                        
\usepackage{nameref}
\usepackage{hyperref}
\usepackage{iris}
\usepackage{aneris}
\usepackage{includes}
\usepackage[disable]{todonotes}
\usepackage[inline]{enumitem}
\usepackage{amsmath}
\usepackage{nameref}
\usepackage{wrapfig}
\usepackage{array}
\usepackage{mathrsfs}
\usepackage{xspace}
\usepackage{pict2e}
\usepackage{tikz}
\usepackage{cleveref}
\usepackage{fontawesome}
\usepackage{graphicx}
\usetikzlibrary{positioning, calc}
\usepackage{caption}
\usepackage{afterpage}
\usepackage{placeins}

\definecolor{darkblue}{rgb}{0.0, 0.0, 0.55}
\definecolor{darkgreen}{rgb}{0.0, 0.55, 0}
\definecolor{darkred}{rgb}{0.55, 0, 0}

\makeatletter
\newcommand{\hourglass}{%
  \mathbin{
  \mspace{1mu}
  \mathpalette\vinc@hourglass\relax}
  \mspace{1mu}
}
\newcommand{\vinc@hourglass}[2]{%
  \begingroup
  \settowidth{\unitlength}{$\m@th#1\mspace{6mu}$}
  \begin{picture}(1,1.732)
  \vinc@linethickness{#1}
  \roundcap\roundjoin
  \Line(0,0)(1,1.732)(0,1.732)(1,0)(0,0)
  \end{picture}%
  \endgroup
}
\newcommand{\vinc@linethickness}[1]{%
  \linethickness{%
      \ifx#1\displaystyle 0.8\fontdimen8\textfont3\else
      \ifx#1\textstyle 0.8\fontdimen8\textfont3\else
      \ifx#1\scriptstyle0.8\fontdimen8\scriptfont3\else
      1\fontdimen8\scriptscriptfont3\fi\fi\fi
  }%
}
\makeatother

\newcommand{\SIrd}{\mathsf{read}}
\newcommand{\SIrdC}[2]{\SIrd ~ #1 ~ #2}
\newcommand{\SIwr}{\mathsf{write}}
\newcommand{\SIwrC}[3]{\SIwr ~ #1 ~ #2 ~ #3}
\newcommand{\SIloop}{\mathsf{loop}}
\newcommand{\SIincl}{\mathsf{init\_client}}
\newcommand{\SIinclC}[1]{\SIincl ~ #1}
\newcommand{\SIinsrv}{\mathsf{init\_srv}}
\newcommand{\SIstart}{\mathsf{start}}
\newcommand{\SIstartC}[1]{\SIstart ~ #1}
\newcommand{\SIcommit}{\mathsf{commit}}
\newcommand{\SIcommitC}[1]{\SIcommit ~ #1}

\newcommand{\SIassert}[1]{\mathsf{assert}(#1)}
\newcommand{\SIletin}[2]{#1 \mathrel{=} #2}
\newcommand{\SIifthen}[1]{\mathsf{If} ~ #1 ~ \mathsf{then}}
\newcommand{\SIcond}{\mathsf{cond}}

\newcommand{\SIwait}{\mathsf{wait}}
\newcommand{\SIwaitC}[3]{\SIwait ~ #1 ~ #2 ~ #3}
\newcommand{\SIrun}{\mathsf{run}}
\newcommand{\SIrunC}[2]{\SIrun ~ #1 ~ #2}

\definecolor{commentcolor}{rgb}{0.0, 0.6, 0.18}

\newcommand{\SeenSymb}{\mathsf{Seen}}
\newcommand{\Seen}[2]{\SeenSymb(#1, #2)}
\newcommand{\mapstoMemSymb}{\mapsto}
\newcommand{\mapstoMem}[2]{#1 \mapstoMemSymb #2}
\newcommand{\mapstoCacheSymb}[1]{\mapsto_{#1}}
\newcommand{\mapstoCache}[3]{#1 \mapstoCacheSymb{#2} #3}
\newcommand{\KeyUpdStatusSymb}{\mathsf{KeyUpdStatus}}
\newcommand{\KeyUpdStatus}[3]{\KeyUpdStatusSymb(#1, #2, #3)}
\newcommand{\ConnectionStateSymb}{\mathsf{ConnectionState}}
\newcommand{\CanStartSymb}{\mathsf{CanStart}}
\newcommand{\CanStart}[1]{\ConnectionStateSymb(#1, \CanStartSymb)}
\newcommand{\ActiveSymb}{\mathsf{Active}}
\newcommand{\Active}[2]{\ConnectionStateSymb(#1, \ActiveSymb(#2))}

\newcommand{\Ppred}[1]{P(#1)}

\newcommand{\CutC}[3]{Cut(#1, #2, #3)}
\newcommand{\Timestamp}{t}

\newcommand{\ValMath}{v}

\newcommand{\KVSinit}{\mathsf{InitKVS}}
\newcommand{\readHandler}{read\_handler}
\newcommand{\readHandlerC}[2]{\readHandler ~ #1 ~ #2}
\newcommand{\TimeFrag}[1]{\mathsf{TimeFrag}(#1)}
\newcommand{\TimeGlobal}[1]{\mathsf{TimeGlobal}(#1)}
\newcommand{\TimeLocal}[1]{\mathsf{TimeLocal}(#1)}
\newcommand{\SnapshotsFrag}[2]{\mathsf{SnapFrag}(#1, #2)}
\newcommand{\SnapshotsAuth}[1]{\mathsf{SnapAuth}(#1)}
\newcommand{\MemGlobal}[1]{\mathsf{MemGlobal}(#1)}
\newcommand{\MemLocal}[1]{\mathsf{MemLocal}(#1)}
\newcommand{\isMap}[2]{\mathsf{isMap} (#1, #2)}
\newcommand{\Model}[3]{\mathsf{Model} (#1, #2, #3)}
\newcommand{\Snapshots}{S}
\newcommand{\Memory}{M}
\newcommand{\AuthSet}[1]{\mathsf{AuthSet}(#1)}
\newcommand{\FragSet}[1]{\mathsf{FragSet}(#1)}

\newcommand{\HistValSymb}{{last}}
\newcommand{\HistVal}[1]{\HistValSymb(#1)}
\newcommand{\HistEmpty}{[~]}
\newcommand{\HistNonEmpty}[1]{[#1]}
\newcommand{\CanCommitSymb}{\mathit{can\_commit}}
\newcommand{\CanCommit}[3]{\CanCommitSymb(#1, #2, #3)}
\newcommand{\CanCommitPredicateSymb}{\mathsf{CanCommit}}
\newcommand{\CanCommitPredicate}[3]{\CanCommitPredicateSymb(#1, #2, #3)}
\newcommand{\CommitHistSymb}{{update\_hist}}
\newcommand{\CommitHist}[2]{\CommitHistSymb(#1, #2)}
\newcommand{\CommitValsSymb}{{update\_vals}}
\newcommand{\CommitVals}[2]{\CommitValsSymb(#1, #2)}
\newcommand{\MapLookup}[2]{#1[#2]}
\newcommand{\UpdateKVSSymb}{\mathit{update\_kvs}}
\newcommand{\UpdateKVS}[3]{\UpdateKVSSymb(#1, #2, #3)}

\newcommand{\Cst}{c}
\newcommand{\Key}{k}
\newcommand{\Keys}{\mathsf{Keys}}
\newcommand{\Hist}{h}
\newcommand{\SetVal}{S}
\newcommand{\Vals}{V}
\newcommand{\HistBig}{H}

\newcommand{\map}{m}
\newcommand{\kvs}{kvs}
\newcommand{\Snapshot}{ms}
\newcommand{\Cache}{mc}
\newcommand{\CacheMath}{C}
\newcommand{\CacheClient}{cache}
\newcommand{\StateClient}{state}
\newcommand{\ValueOption}{ov}
\newcommand{\Boolean}{b}

\newcommand{\Body}{body}
\newcommand{\srvsa}{srv}

\newcommand{\RU}{\scaleto{\mathsf{RU}}{3.5pt}.}
\newcommand{\RC}{\scaleto{\mathsf{RC}}{3.5pt}.}
\newcommand{\SI}{\scaleto{\mathsf{SI}}{3.5pt}.}

\newcommand{\GlobalInvBody}{\mathsf{GlobalInv}}

\newcommand{\GlobalInv}{\knowInv{}{\GlobalInvBody}}

\newcommand{\FinishToken}{\mathsf{T}}

\allowdisplaybreaks

\begin{document}

\title{Reasoning about Weak Isolation Levels in Separation Logic}

\author{Anders Alnor Mathiasen}
\orcid{0009-0005-6587-5590}
\affiliation{%
  \institution{Aarhus University}
  \city{Aarhus}
  \country{Denmark}
}
\email{alnor@cs.au.dk}

\author{Léon Gondelman}
\orcid{0000-0001-8262-6397}
\affiliation{%
  \institution{Aalborg University}
  \city{Aalborg}
  \country{Denmark}
}
\email{lego@cs.aau.dk}

\author{Léon Ducruet}
\orcid{0009-0002-6141-5624}
\affiliation{%
  \institution{Aarhus University}
  \city{Aarhus}
  \country{Denmark}
}
\email{leon.ducruet@ens-lyon.fr}

\author{Amin Timany}
\orcid{0000-0002-2237-851X}
\affiliation{%
  \institution{Aarhus University}
  \city{Aarhus}
  \country{Denmark}
}
\email{timany@cs.au.dk}

\author{Lars Birkedal}
\orcid{0000-0003-1320-0098}
\affiliation{%
  \institution{Aarhus University}
  \city{Aarhus}
  \country{Denmark}
}
\email{birkedal@cs.au.dk}

\begin{abstract}
  Consistency guarantees among concurrently executing transactions in local- and distributed systems,
  commonly referred to as isolation levels, have been formalized in a number of models.  
  Thus far, no model can reason about executable implementations of databases or local transaction libraries providing weak isolation levels.
  Weak isolation levels are characterized by being highly concurrent and, unlike their stronger counterpart serializability, they are not equivalent to the consistency guarantees provided by a transaction library implemented using a global lock.
  Industrial-strength databases almost exclusively implement weak isolation levels as their default level. This calls for formalism as numerous bugs  
  violating isolation have been detected in these databases. \\
  \indent In this paper, we formalize three weak isolation levels in separation logic, namely \emph{read uncommitted}, \emph{read committed}, and \emph{snapshot isolation}.
  We define modular separation logic specifications that are independent of the underlying transaction library implementation.
  Historically, isolation levels have been specified using examples of executions between concurrent transactions that are not allowed to occur, and we demonstrate that our specifications correctly prohibit such examples.
  To show that our specifications are realizable, we formally verify that an executable implementation of a key-value database running the multi-version concurrency control algorithm from the original snapshot isolation paper satisfies our specification of snapshot isolation.
  Moreover, we prove implications between the specifications---snapshot isolation implies read committed and read committed implies read uncommitted---and thus the verification effort of the database serves as proof that all of our specifications are realizable.
  All results are mechanized in the Rocq proof assistant on top of the Iris separation logic framework.
\end{abstract}




\maketitle


\section{Introduction}
\label{sec:1}

Transactions, defined as a set of operations considered as a single unit, 
is a common concept implemented by most modern databases and can also be encountered in local transactional memory systems.
There exists multiple isolation levels for describing the allowed interaction between concurrently executing transactions. 
The strong isolation level of serializability \citep{orig-serializability} often serves as an introduction to isolation levels. 
This level has consistency guarantees  equivalent to what one would achieve in a transactional library protected 
by a global lock acquired at start time and released at commit time by the transactions. 
While serializability can be a useful property, it is often too strong for the desired use case and will negatively impact the 
throughput of the system. In contrast to serializability, weaker isolation levels allow for more concurrency and higher throughput. 
In fact, most modern commercial database systems, e.g., MySQL, Microsoft SQL, Oracle and PostgreSQL, have a weaker isolation level than serializability as default
(Microsoft SQL, Oracle and PostgreSQL uses read committed \citep{postgresql, oracle,microsoftsql} which we specify in this paper). 
With increased concurrency comes weaker consistency guarantees that by nature are harder to capture and reason about. 
As numerous bugs have been found in commercial database implementations violating isolation \citep{jepsen}, 
it is highly important to formally verify database implementations and their clients.

\paragraph{Existing Formalizations of Isolation Levels} A number of different formalizations have been developed to capture and 
reason about isolation levels. Historically, in the ANSI SQL standard of 1992, isolation levels were defined using examples 
of executions between concurrently executing transactions, so-called phenomena, which must 
be prohibited by the particular isolation level. The isolation levels of read uncommitted, read committed, repeatable read and 
serializability were defined using a phenomenon for each level except read uncommitted. This was later critiqued for being imprecise, which led to the invention of 
snapshot isolation and improved phenomena by \citet{orig-SI}. With the phenomena as a basis, a number of different 
models have been developed to reason about isolation levels. Predominantly, models have been based on dependency graphs \citep{Adya99, Adya00},
operational semantics \citep{Xiong2020, transactions-gotsman, Kaki17, Crooks17} or abstract executions \citep{Burckhardt12, Ketsman23}.
While the mentioned models can be used to define the desired consistency guarantees for transactions, in the form 
of isolation levels, they can not be applied to verify executable implementations of databases,
transactional libraries and clients thereof. 
Recently, the separation logic framework Iris \citep{irisjournal}
has been used in \citet{vMVCC} to give a modular specification 
for serializability in the context of a key-value store. 
Separation logic is well known for its ability to specify various properties, using Hoare triples, 
and being able to reason about executable implementations 
\citep{DBLP:journals/pacmpl/GondelmanGNTB21,nieto_et_al:LIPIcs.ECOOP.2023.22,abel/crdts,grove}.
The specification for serializability in \cite{vMVCC} leverages that serializability is a strong property: 
A concise specification is given for a run operation that bundles together the start operation, commit operation and 
intermediate operations of a transaction into a single specification --- something which is unfeasible 
for weaker isolation levels (this is discussed in \Cref{sec:8}).
The argument of using separation logic to capture isolation levels and reason about executable code
has yet to be completed by showing that separation logic can capture weak isolation levels. 

\paragraph{Specifying Weak Isolation Levels in Separation Logic} 
The main contributions of this paper are formal separation logic specifications
of weak isolation levels: read uncommitted, read committed and snapshot isolation.
Together with serializability and repeatable read, these levels constitute the common levels found in 
commercial systems (repeatable read is identical to serializability except for its behavior regarding the 
predicate-read operation \citep{Adya00} which we do not model).
To demonstrate that our specifications capture the desired isolation levels, 
we have, for the phenomena in the literature, created and proven examples of
client transactions running against libraries adhering to our specifications which show that the phenomena 
are prohibited. 
We have chosen to verify the phenomena from the literature as clients because 
(1) They are concise examples that exercises edge cases, which a larger client implementation
may not necessarily cover; and 
(2) these client-examples are used as the only specification in the SQL standard, 
and this is what industry databases use as specification. 
In addition to the phenomena, we have also verified a bank transfer example 
that represented a common use case for transactions (\Cref{sec:5}). 
Furthermore, we prove implications between our specifications, snapshot isolation implies 
read committed and read committed implies read uncommitted, 
which is consistent with the literature \citep{Adya99}. 
Implications between separation logic specifications are unusual, as they establish that any 
implementation of the assumed specification also implements the implied specification.

\paragraph{Realizing Specifications} To show that our specifications are realizable 
by concrete implementations, 
we implement a database and verify that it meets 
our snapshot isolation specification. Together with the implication proofs between
the specifications, this serve as a proof that all of our specifications are realizable.
In more detail, we have used the distributed separation logic of Aneris 
\citep{DBLP:conf/esop/Krogh-Jespersen20} 
to formally verify an in-memory single-node key-value store with support for transactions. 
In fact, we have implemented and verified the algorithm from the original 
snapshot isolation paper by \citet{orig-SI}.
The implementation uses multi-version concurrency control where for each key all 
previous updates are stored and ordered by timestamps.
In addition to the operations in the API of the database, we have implemented utility 
functionalities on top of these operations in the form of wait and run operations.
The wait operation returns once it has observed the key-value pair it has been given as argument, 
while the run operation wraps start and commit operations around a transaction body as in \citet{vMVCC}.
We verify specifications for the wait operation for all the isolation levels we are considering,  
while we argue that only snapshot isolation is strong enough to 
have a concise specification for the run operation.
Our implementations are written in OCaml and transpiled to AnerisLang 
for mechanized verification in the Rocq proof assistant.\footnote{The Coq proof assistant has recently been renamed to Rocq.}
In fact, all the results in this paper have been mechanized in Rocq.
Aneris is an instantiation of the Iris separation logic framework 
with an OCaml like language (AnerisLang) and semantics for unreliable network communication.
We emphasize that while we have used Aneris for the verification and implementation
of a database, our specifications are independent of Aneris 
and can be used with other instantiations of Iris/separation logics. 

\paragraph{Challenges} One of the key challenges of this work is to capture the isolation levels by modular 
separation logic specifications, with formal specifications for each of the individual operations 
on the key-value store --- traditionally, isolation levels have not been formulated 
at this level of granularity with separate specifications for each of the operations: 
Existing transactional models either have a global view of all transactions, or are formulated with 
a high-level idealized operational semantics far from the semantics used to model the programming language on which 
we build in this paper. We emphasize that one of the advantages of our approach, using separation logic specifications, is modularity in 
because it enables one to reason formally about client programs that (possibly among other libraries) make use of the key-value store as a library. 
Another key challenge is to prove the implications between the specifications of the three isolation 
levels. The specifications are highly modular as they keep the logical resources abstract by hiding the definitions. 
In the implication proofs, we must construct separation logic resources 
used in the implied specification, given the abstract separation logic resources of the assumed specification,  
in a way that lets us show the formal specifications for each of the individual operations of the implied specification.
That is while hiding the fact that the newly constructed logical resources are defined in terms of the assumed logical resources.
Lastly, to verify that the multi-version concurrency control algorithm of \citet{orig-SI} 
implements our separation logic specifications, in the context of a distributed system, requires 
non-trivial and substantial proof effort (more than 5000 lines of Rocq proof code), especially since our specification for snapshot isolation 
includes evidence for unsuccessful commits in the form of a conflict between concurrent transactions
expressed as a resource in the postcondition of the commit specification.

\paragraph{Contributions} In summary, our paper makes the following contributions:
\begin{enumerate}
\item The first modular separation logic specifications for weak isolation levels, in particular  
      read uncommitted, read committed and snapshot isolation.
      The specifications are independent of the underlying algorithm and whether the system is a 
      local or distributed system, and we provide examples that show the specifications prohibit the 
      phenomena of the literature. 
\item Formal proofs showing that the specification of snapshot isolation implies the specification of read committed, and that the specification of read committed implies the specification of read uncommitted.
\item Implementation and verification of an in-memory single-node key-value database with snapshot isolation.
      The implementation is based on the original multi-version concurrency control snapshot isolation algorithm, 
      and is the first formally verified executable database supporting transactions.
\end{enumerate}
This gives us, for the first time, one unified logic in which we can verify clients and database implementations for 
transactions with weak isolation levels. 
That is all while enjoying the modularity of higher-order separation logic such as the ability to
combine code using transactions with other verified libraries (something which existing work do not address).
All results in this paper are mechanized in the Rocq proof assistant using the Iris separation logic.

\paragraph{Structure of the paper} The rest of this paper is structured as follows: 
In \Cref{sec:2}, we present our specification for read uncommitted using examples 
we have verified. \Cref{sec:2.2} 
considers the proof of a particular example in detail using the read uncommitted specification.
In \Cref{sec:3}, we present our specification for read committed and
the examples we have proven using the read committed specification.
In \Cref{sec:4}, we prove the specification for read committed implies the specification 
for read uncommitted. 
In \Cref{sec:5}, we present our specification for snapshot isolation. 
Here, we also present the examples we have proven using the snapshot isolation specification.
In \Cref{sec:6}, we prove the specification for snapshot isolation implies the specification 
for read committed. 
In \Cref{sec:7}, we explain the verification of our database implementation with 
respect to our specification for snapshot isolation.
In \Cref{sec:8} we discuss related work, and in \Cref{sec:9} we talk about future work before concluding. 
We will refer to \Cref{sec:app_util} for presentation of a 
utility wait operation for all the isolation levels and a utility 
run operation for snapshot isolation together with a discussion of why 
a run operation is impractical for read uncommitted and read committed.


\section{Specifying Read Uncommitted}
\label{sec:2}

In this section, we first recall the informal description of read uncommitted, 
and then present our modular separation logic specification in \Cref{sec:2.1}.
This is done using the so-called \emph{read uncommitted data} and 
\emph{read own data} examples which we have verified using our specifications. 
Afterwards, in \Cref{sec:2.2}, we show in some detail how we can use the specification 
to prove safety and correctness of the concurrently executing transactions in the 
\emph{read uncommitted data} example.
We present the proof to help the reader get acquainted with using the style of specifications 
presented throughout the paper.

\paragraph{Read uncommitted} Read uncommitted is the weakest isolation level.
The level has mostly theoretical value, and it is often used in transactional models to establish a baseline, of definitions
and structure, which we can use to define stronger levels. We follow the same approach.
The only guarantees one gets from using read uncommitted is that data being read has been written by some transaction
(captured by the \emph{read uncommitted data} example in \Cref{fig:read-uncommitted-example})
, and if a read comes after a write from the same transaction, then the read will be reading from this write 
(captured by the \emph{read own data} example in \Cref{fig:read-own-example}).

The \emph{read uncommitted data} example 
(we ignore the invariant in \color{darkblue} blue \color{black} for now)
consists of two transactions (separated by the vertical bars), which are executed concurrently on two different nodes in a distributed system, alternatively threads in a local system, in which there is also a server with the key-value store.
Notice that the example includes an assert statement, which crashes if the expression within it does not hold.
We use assert because it makes it easy to state (in a logic-independent way) that an intended property holds.
In this example, the assert statement expresses that the reading transaction will read nothing or the value written by the writing transaction, even though the latter does not commit.
This is possible in read uncommitted as there is no guarantee that transactions will read committed data; it is only guaranteed that the data has been written by another transaction --- independently of whether this transaction aborted or committed.

The \emph{read own data} example follows a similar structure with two transactions. 
Here the assert statement expressed that no matter when the write of the first transactions is scheduled, the second 
transactions will always read the value previously written by itself. 
\begin{figure}[h]
  \small
  \centering
  \vspace{-0.3cm}
    \begin{minipage}[b]{0.5\textwidth}
      \begin{align*}
        \left.
        \begin{aligned}
          & \SIstart\\
          & \SIwrC{x}{1}\\
          & \SIloop\\
          &
        \end{aligned}
        ~\middle\Vert~
        {
        \begin{aligned}
          & \SIstart\\
          & \SIletin{v_x}{\SIrdC{x}}\\
          & \SIassert{v_x = \None \lor v_x = \Some{1}}\\
          & \SIcommit
        \end{aligned}
        }
        \right.
        \end{align*}
        \begin{align*}
          Inv &\eqdef{}  \color{darkblue} \exists V,~ \RU\mapstoMem{x}{V} ~ \ast (V = \emptyset \lor V = \{1\}) 
        \end{align*}  
    \vspace{-0.4cm}
    \caption{Read uncommitted data.}
    \label{fig:read-uncommitted-example}
  \end{minipage}%
  \begin{minipage}[b]{0.5\textwidth}
    \begin{align*}
      \left.
      \begin{aligned}
        & \SIstart\\
        & \SIwrC{x}{1}\\
        & \SIcommit\\
        &
        &
      \end{aligned}
      ~\middle\Vert~
      {
      \begin{aligned}
        & \SIstart\\
        & \SIwrC{x}{2}\\
        & \SIletin{v_x}{\SIrdC{x}}\\
        & \SIassert{v_x = \Some{2}}\\
        & \SIcommit
      \end{aligned}
      }
      \right.
      \end{align*}
    \caption{Read own data.}
    \label{fig:read-own-example}
  \end{minipage}
  \vspace{-0.7cm}
\end{figure}

Some formalizations of read uncommitted, e.g., \citet{Adya99}, imposes restrictions on the 
order in which versions are installed inside the key-value store. As separation logic abstracts away 
from implementation details, this is not a requirement we can or want to impose. 
This choice is in line with the work of \citet{Crooks17} which takes a client centric approach.

\subsection{Separation Logic Specifications}
\label{sec:2.1}

Our formal specifications of the read uncommitted operations are defined using the Aneris program logic 
\citep{DBLP:conf/esop/Krogh-Jespersen20}.
Aneris is a higher-order distributed separation logic, based on the Iris framework \citep{irisjournal}, 
fully mechanized in the Rocq proof assistant.\footnote{While we happened to use Aneris, we could have used another distributed 
separation logic based on the Iris framework such as Grove \citep{grove}, see the discussion in \Cref{sec:8}.} 
Our specifications are written using Hoare triples 
$\anhoare{P}{\expr}{\Ret \val. Q}{\ip}{}$.
A Hoare triple asserts that given a machine state satisfying the precondition $\prop$, 
then the execution of the expression $\expr$ is safe and, moreover, if the execution terminates 
then $Q$ holds for the resulting machine state when the return value is bound to $\val$. 
In Aneris, Hoare triples are annotated with the IP-address of the node on which the expression 
is executing, but we often omit this annotation when the IP-address is not of importance or for presentation purposes.
In fact, for most parts of this paper (except for the proof of the snapshot isolation implementation in \Cref{sec:7}),
one can just pretend that the logic is the standard Iris logic; no special knowledge of Aneris is required
to understand the paper.
The propositions in the precondition and postcondition of the Hoare triples are often referred 
to as resources, since Aneris is a separation logic with a concept of ownership.
We recall that the separating conjunction $\prop \ast Q$ only can be proven from a resource that 
can be split into two resources satisfying $\prop$ and $Q$ respectively.
The magic wand $\prop \wand Q$ is closely related to the separating conjunction: 
in combination with exclusive ownership of $\prop$, then $\prop \wand Q$ entails $Q$.
Some resources denote pure knowledge and do not assert exclusive ownership; this holds in 
particular for all persistent predicates, which are those satisfying that $\always \prop \vdash \prop$, 
where $\always$ is the so-called persistence modality which, intuitively, takes a predicate and 
forgets about the resources exclusively owned by the predicate.
Persistent predicates can thus always be freely duplicated.
For instance, Hoare triples are defined using the persistent modality, 
to allow us to use our specifications an unbounded number of times.

\begin{figure}
  \centering
  \small
  \color{darkblue}
  \begin{mathpar}
    \axiomH{ru-init-client-spec}
    { 
      \anhoare
        {
          \TRUE
        }
        {
          \SIinclC{\srvsa}
        } 
        { \begin{array}[t]{@{}l@{}}
          \Ret \Cst. \RU\CanStart{\Cst}
          \end{array}
          }
        {}{}
    }
    \and
    \axiomH{ru-init-kvs-spec}
    { 
      \anhoare
        { \begin{array}[t]{@{}l@{}}
          \RU\KVSinit
          \end{array}
        }
        { 
          \SIinsrv
        } 
        { \begin{array}[t]{@{}l@{}}
          \Ret \TT. \TRUE
        \end{array}}
        {\srvsa}{}
    }
    \and
    \axiomH{ru-start-spec}
    { 
      \anhoareVatomic
        { \begin{array}[t]{@{}l@{}}
          \RU\CanStart{\Cst} \ast
          \Sep_{(\Key,\, \Vals) \in\, \map} \RU\mapstoMem{\Key}{\Vals} 
        \end{array}
        }
        { \begin{array}[t]{@{}l@{}}
          \SIstartC{\Cst}
        \end{array}
        } 
        { \begin{array}[t]{@{}l@{}}
          \Ret \TT. \RU\Active{\Cst}{\dom(\map)} \asts \\
          \Sep_{(\Key,\, \Vals) \in\, \map} \Big(\RU\mapstoMem{\Key}{\Vals}\ \asts
          \RU\mapstoCache{\Key}{\Cst}{\None}\Big)
        \end{array}}
        {}{}
    }
    \and
    \axiomH{ru-read-spec}
    { 
      \anhoareVatomic
        { \begin{array}[t]{@{}l@{}}
          \RU\mapstoCache{\Key}{\Cst}{\ValueOption} \ast \RU\mapstoMem{\Key}{\Vals}
          \end{array}
        }
        { \begin{array}[t]{@{}l@{}}
          \SIrdC{\Cst}{\Key}
          \end{array}
        } 
        { \begin{array}[t]{@{}l@{}}
          \Ret ow.
          \RU\mapstoCache{\Key}{\Cst}{\ValueOption} \ast \RU\mapstoMem{\Key}{\Vals} \asts \\
          \Big(\big(\ValueOption = \None ~ \land \\
          \quad ((\exists v, \mathit{ow} = \Some{v} \land v \in \Vals) ~ \lor \\
          \qquad ow = \None)\big) ~ \lor \\ 
          \big(\ValueOption \neq \None \land ow = \ValueOption \big)\Big)
        \end{array}}
        {}{}
    }
    \and
    \axiomH{ru-commit-spec}
    { 
      \anhoareVatomic
        { \begin{array}[t]{@{}l@{}}
          \Active{\Cst}{\SetVal} \ast
          \SetVal = \dom ~ \map =  \dom ~ \Cache \asts \\[0.4em]
          \Sep_{(\Key,\, \Vals) \in\, \map} \RU\mapstoMem{\Key}{\Vals} ~ \asts
          \Sep_{(\Key,\, \ValueOption) \in\, \Cache} \RU\mapstoCache{\Key}{\Cst}{\ValueOption}
          \end{array}
        }
        { \begin{array}[t]{@{}l@{}}
          \SIcommitC{\Cst}
        \end{array}
        }
        { \begin{array}[t]{@{}l@{}}
          \Ret \val. \RU\CanStart{\Cst} \ast \Sep_{(\Key,\, \Vals) \in\, \map} \RU\mapstoMem{\Key}{\Vals}
        \end{array}
        }
        {}{}
    }
    \and
    \axiomH{ru-write-spec}
    { 
      \anhoareVatomic
        {
         \begin{array}[t]{@{}l@{}}
          \RU\mapstoCache{\Key}{\Cst}{\ValueOption} \ast \RU\mapstoMem{\Key}{\Vals}
        \end{array}
        }
        {
          \SIwrC{\Cst}{\Key}{\val}
        } 
        { \begin{array}[t]{@{}l@{}}
          \Ret \TT. \RU\mapstoCache{\Key}{\Cst}{\Some \val} \asts \\ \RU\mapstoMem{\Key}{(\Vals \cup \{\val\})}
          \end{array}
          }
        {}{}
    }
    \and
    \end{mathpar}
    \caption{Specification for read uncommitted ($\color{darkblue}\mathsf{RUSpecs}$).}
    \label{fig:specs_ru}
    \vspace{-0.5cm}
  \end{figure}
We now begin to describe the specifications of the operations of read uncommitted, 
which can all be found in \Cref{fig:specs_ru}. 
We motivate and introduce the concepts used in the specifications along the way.
For notation purposes, as we will be presenting specifications
for other isolation levels with similar notation, we use the color \color{darkblue} blue \color{black}
and the prefix $\color{darkblue}\RU$ when presenting the read uncommitted specification. 
Later, we will use similar notation for the other isolation levels.

\paragraph{Init client operation} The specifications for all our isolation levels are 
centered around exclusive points-to resources.
This is a popular choice in separation logic as it allows specifications only to require 
knowledge of a single key/location as opposed to a larger state.
We will be working with two types of exclusive points-to resources: 
A \emph{local points-to resource} per active transaction, denoted 
$\color{darkblue}\RU\mapstoCache{\Key}{\Cst}{\ValueOption}$ for read uncommitted, which
asserts that the key $\Key$ points to an optional value $\color{darkblue}\ValueOption$ for the transaction 
active on connection $\Cst$, and a \emph{global points-to resource}, denoted 
$\color{darkblue}\RU\mapstoMem{\Key}{\Vals}$ which asserts that the key $\color{darkblue}\Key$ points to 
a state consisting of a set of values $\Vals$, 
which captures the (global) state of the key-value store.
A client only gets to make requests to the key-value store once
a connection has been established with the provider of the key-value store, whether it is a local 
connection or happens over the network.
A connection can be established by the client calling the $\SIinclC$ method, 
whose specification \ruleref{ru-init-client-spec}
is in \Cref{fig:specs_ru}.
A client can always try to establish a connection with the key-value store 
(the precondition is simply $\TRUE$), and
in the postcondition, the client gains proof that it has a connection in the form of 
the resource $\color{darkblue}\CanStart{\Cst}$.
The $\SIinclC$ method keeps trying to establish a connection 
to the server with IP-address $srv$ when called, thus if no key-value store 
is available on the IP-address, 
the call will not terminate. 
If a client has multiple connections to different key-value stores, 
the points to resources will also take as argument the IP-address 
of the key-value store to which they belong. 
As long as there is only one active key-value store, the IP-address 
argument is implicit.

\paragraph{Start operation} As potentially many clients concurrently want to start their transactions, the resources capturing the key-value store state will often reside in an invariant.
Invariants are propositions that hold at every step of execution once they have been established.
An invariant holding a resource $\prop$ is denoted $\knowInv{}{\prop}$.
\footnote{A reader familiar with Iris will notice that we have omitted invariant names, and in later 
parts of the paper masks, to make the presentation less technical.}
In Iris, and hereby Aneris, one can only open invariants across atomic expressions (expressions which in the operational semantics can step to a value using a single step).
For this reason, we use \emph{logical atomic triples} instead of standard Hoare triples to specify the transaction operations (for notation, logical atomic triples uses angle brackets as opposed to the curly brackets of ordinary Hoare triples).
We refer to \citet{iris} for a detailed description of logical atomic triples; in this paper, we simply treat them as special Hoare triples that allow us to open invariants around them.
In \Cref{fig:inv_atomic}, we display the rule used to open an invariant 
around atomic Hoare triples. 

\begin{figure}[!h]
  \vspace{-0.2cm}
  \centering
  \begin{mathpar}
  \inferH{Inv-atomic}
  {\proves \anhoareatomic{R \ast P}{\expr}{\Ret \val. Q \ast R}{}{}}
  {\knowInv{}{R} \proves \anhoareatomic{P}{\expr}{\Ret \val. Q}{}{}}
  \end{mathpar}
  \caption{Invariant rule for atomic Hoare Triples.}
  \label{fig:inv_atomic}
  \vspace{-0.2cm}
\end{figure}

\noindent With the rule \ruleref{Inv-atomic}, we can use the resources in an invariant to 
prove the precondition of an atomic Hoare Triple, 
and we can use the resources in the postcondition 
of the atomic Hoare Triple to reestablish the invariant.

Now, the \ruleref{ru-start-spec} rule takes as precondition the connection state together with global 
points-to resources for all the keys on which the transaction wants to perform 
reads and writes -- these are the keys in the map $\color{darkblue}\map$ in the notation.
As the key-value store state, in the form of global points-to 
resources, can be shared among many clients in an invariant, the start operation
is specified using a logical atomic triple. The postcondition of the start operation states 
that the connection is active for the keys in the domain of $\color{darkblue}\map$
(this information will be used later at commit time) and gives back the global 
points-to resources together with fresh local points-to resources.\footnote{Separation-logic aficionados who are familiar with the frame rule may wonder why the global points-to resources are included unchanged in both the precondition and the postcondition; we have decided to include them (both for start and commit) in order to be able to prove formally that read committed implies read uncommitted, see \Cref{sec:4}.}

Now, we could have elided an explicit $\SIinclC$ operation and made it part of the $\SIstart$
operation to simplify the specification. 
This would, however, imply initialization overhead 
(cache and network-connection creation) on a per-transaction basis which is suboptimal. 

\paragraph{Read and write operation} Having started a transaction with the start operation, 
an active transaction can perform read and write operations using its newly gained 
local points-to resources.
The read operation requires the global and local points-to resources for the key in its precondition. 
If the active transaction has already written to the key, this value will be read straight 
from the local points-to resource. Otherwise, the value being read can be $\color{darkblue}\None$
or come from the global state, i.e., the value will be from the set of values $\color{darkblue}\Vals$ pointed 
to by the global points-to resource. 
As for the read operation, the precondition of the specification for the write operation requires 
global and local points-to resources for the key that is being written to. 
As we are specifying read uncommitted where transactions are allowed to read uncommitted data, 
both the global and local points-to resource is updated. 
Note, in particular, that the global points-to resource gets updated immediately
for other transactions to read, even if the writing transaction does not commit.

At this point, it can be tempting to create a stronger specification. 
For instance, a meaningful optimization would be to 
cache a local read and
update $\mathit{ov}$ to $\mathit{ow}$ when $\mathit{ov}$ is none for local points-to resource 
in the read specification.
But if we enforce such an optimization in the specification, 
then we are specifying something stronger than read uncommitted 
(not all implementations of read uncommitted have this optimization). 
Do notice that any implementation doing the optimization can implement our specification, 
it will simply pick $\mathit{ow}$ from the set $\Vals$ on a subsequent read
if $\mathit{ov}$ remains $\None$.

\paragraph{Commit operation} At commit time, the connection state, local and global points to 
resources are collected in the precondition of the \ruleref{ru-commit-spec} rule. 
The domain conditions ensure that local and global points-to resources are 
available for all the keys on which the transaction was started.
No matter if the commit is successful or not (the return value 
of commit is a boolean), 
the state of the key-value store remains the same in the postcondition of the 
\ruleref{ru-commit-spec}, as all the updates made by the transaction 
have already been propagated at the time of the write operation, cf. \ruleref{ru-write-spec}, 
to the global points-to resources. Thus, the job of the commit specification in 
read uncommitted is simply to collect the local points-to resources and change the connection state.

\paragraph{Init KVS operation and the complete specification} 
Initializing the key-value store with \ruleref{ru-init-kvs-spec}
requires the $\color{darkblue}\RU\KVSinit$ resources which can not be obtained 
from using the per operation specifications in \Cref{fig:specs_ru}. 
Likewise, initial global points to resources 
must be obtained before one can start using the specifications in \Cref{fig:specs_ru}.
Thus, a complete specification for an isolation level must provide 
all the resources we need in order to get started with using 
the per operation specifications. Below, in \eqref{eq:ru-setup_specs}, 
we have defined exactly what the complete 
read uncommitted specification $\color{darkblue}\mathsf{RUSpec}$ is. 
The complete specification $\color{darkblue}\mathsf{RUSpec}$ 
is stated using the Iris update modality.
The Iris update modality $\pvs[][] \prop$ states that 
initialization of resources or updates to currently owned resources
can be made to satisfy $\prop$.
In the context of $\color{darkblue}\mathsf{RUSpec}$,
this means we can initialize the resources 
that constitute the specification.
We will continue with explaining each of the 
components $\color{darkblue}\mathsf{RUSpec}$ consists of. 

\begin{equation}
  \small
  \color{darkblue}
  \mathsf{RUSpec} \eqdef 
    \pvs[]{} ~ 
    \RU\KVSinit \ast
    \Sep_{\Key\, \in\, \Keys} \RU\mapstoMem{\Key}{\emptyset} \ast
    \mathsf{RUSpecs} \ast
    \RU\GlobalInv 
  \color{black}
\label{eq:ru-setup_specs}
\end{equation}

First, we have the resource $\color{darkblue}\RU\KVSinit$ to initialize the key-value store
as mentioned. Next, for every key we have $\color{darkblue}\RU\mapstoMem{\Key}{\emptyset}$ with each key 
pointing to the empty set as the initial state.
$\color{darkblue}\mathsf{RUSpecs}$ is defined as all the specifications from \Cref{fig:specs_ru}, concretely 
it is the separated conjunction between each of the Hoare Triples in \Cref{fig:specs_ru}.
Lastly, we have a global invariant $\color{darkblue}\RU\GlobalInv$, which we have not yet seen a use case for.
The global invariant is needed to be able to prove two properties which we have 
named the \emph{exclusion} and \emph{creation} properties. We will first motivate and 
then explain these properties next.

\paragraph{The exclusion and creation properties}
As already mentioned, the state of key-value store, in 
the form of global points-to resources, will be shared using an invariant when concurrent access
is needed (see the following subsection for an example). 
The invariant has to be of a form that captures the key-value 
store states that are created by the transactions. Not all states will be valid at any given time, 
for instance, if a value has been observed it can not disappear later. 
To capture such observations, our specifications (for all our isolation levels) come with 
a \emph{seen} resource, which works together with the global points-to resource.
The $\color{darkblue}\Seen{\Key}{\Vals}$ resource for read uncommitted 
intuitively captures the observation of having seen 
the key-value store in the state where
$\color{darkblue}\Key$ was pointing to the set of values $\color{darkblue}\Vals$.
Clients can use this knowledge to exclude absurd states of an invariant.
The exclusion of key-value store states based on observations happens through the 
exclusion property \eqref{eq:ru-exclusion-rule}.

\begin{equation}
  \color{darkblue}
  \small
    \RU\GlobalInv \ast \RU\Seen{\Key}{\Vals'} \ast \RU\mapstoMem{\Key}{\Vals} 
    \proves \pvs[]{} ~ \RU\mapstoMem{\Key}{\Vals} \ast \Vals' \subseteq \Vals
  \color{black}
    \label{eq:ru-exclusion-rule}
\end{equation}
\vspace{-0.3cm}
\begin{equation}
  \color{darkblue}
  \small
    \RU\GlobalInv \ast \RU\mapstoMem{\Key}{\Vals} 
    \proves \pvs[]{} ~ \RU\mapstoMem{\Key}{\Vals} \ast \RU\Seen{\Key}{\Vals}
  \color{black}
    \label{eq:ru-creation-rule}
\end{equation}
\vspace{-0.3cm}

\noindent The essence of the exclusion property is that given an observation 
$\color{darkblue}\RU\Seen{\Key}{\Vals'}$ and the current state of a key
$\color{darkblue}\RU\mapstoMem{\Key}{\Vals}$, 
it must be the case that the current state is the same as the observation or 
the result of having applied updates to the observation,
i.e., $\color{darkblue}\Vals' \subseteq \Vals$ holds. 
The seen resources are created by the creation property \eqref{eq:ru-creation-rule}. 
To be able to use the creation and the exclusion properties we must 
have access to the global invariant of the
read uncommitted key-value library which is therefore a part of the complete 
specification $\color{darkblue} \mathsf{RUSpec}$.
\footnote{The $\SeenSymb$ resources can also be defined as a standalone 
library that will assume the remaining read uncommitted specification.}

We remark that in the exclusion property \eqref{eq:ru-exclusion-rule}
and the creation property \eqref{eq:ru-creation-rule}, the
$\color{darkblue}\RU\GlobalInv$ and $\color{darkblue}\RU\Seen{\Key}{\Vals'}$ 
resources are not included on the right-hand sides but can be found on the left-hand 
sides of the entailment; it is not necessary to include them on the right
since both of these predicates are persistent and hence can be freely
duplicated prior to applying the exclusion property rule. 

\subsection{Proof of Read Uncommitted Data Example}
\label{sec:2.2}

In this section, we show how to use the specifications from \Cref{fig:specs_ru} to prove the \emph{read uncommitted data} example shown in \Cref{fig:read-uncommitted-example}.
What we mean by ``proving the read uncommitted data example'' is that we prove a specification (a Hoare triple) for this example.
By adequacy of the Aneris logic, this then means that the assert shown in \Cref{fig:read-uncommitted-example} will not fail.
In other words, by proving a specification for the example, we show that the property expressed in the assert statement holds for any interleaving.

We begin by using the setup rule \eqref{eq:ru-setup_specs}. 
As the key-space for the example consists only of $x$, the setup rule gives us 
$\color{darkblue}\RU\mapstoMem{x}{\HistEmpty}$. This resource is then used to create the client invariant
$\knowInv{}{\color{darkblue} \exists V,~ \mapstoMem{x}{V} ~ \ast (V = \emptyset \lor V = \{1\})}$, 
also shown in \Cref{fig:read-uncommitted-example} (all the examples in this paper uses invariants 
but we only state them when it is relevant for the presentation).
The $\color{darkblue}\RU\KVSinit$ resource from the setup is used to initialize the key-value store using 
\ruleref{ru-init-kvs-spec}.
We then prove each of the two client transactions separately --- 
this is possible in Aneris as long as one 
distributes resources appropriately among the clients prior to doing the individual proofs. 
As invariants are persistent and thus duplicable resources, we can create two copies of 
$\knowInv{}{\color{darkblue} \exists V,~ \mapstoMem{x}{V} ~ \ast (V = \emptyset \lor V = \{1\})}$
and use one for each of the individual clients proofs.
The sharing of invariants is only safe as long as each client 
owning an invariant uses it safely (this is enforced in Iris). 
Namely, the resources in an invariant can be accessed 
around operations specified with atomic Hoare triples or atmoic expressions where we use the former.
For both transactions it is the case that their proofs are bootstrapped using \ruleref{ru-init-client-spec}, 
which means that in each case, we have a $\color{darkblue}\RU\CanStart{\Cst}$ resource 
for their connection $\color{darkblue}\Cst$ (each client have a unique connection). 
We will now proceed by proving the clients transactions one at a time.

\paragraph{Write-transaction} The transaction is started using the start operation.
To satisfy the precondition of the \ruleref{ru-start-spec} rule, we provide $\color{darkblue}\RU\CanStart{\Cst}$ 
and the global points-to resources from the client invariant, i.e., 
$\color{darkblue}\RU\mapstoMem{x}{\Vals}$ for a fresh set of values $\color{darkblue}\Vals$ for which 
$\color{darkblue} V = \emptyset \lor V = \{1\}$ holds.
By the \ruleref{ru-start-spec} rule we then 
obtain the resources $\color{darkblue} \RU\Active{\Cst}{\{x\}} \ast \RU\mapstoMem{x}{\Vals} 
\ast \RU\mapstoCache{x}{\Cst}{\None}$. 
The global points-to resource $\color{darkblue}\RU\mapstoMem{x}{\Vals}$ is used to close the 
client invariant while we hold onto the rest of the resources.
We then use the \ruleref{ru-write-spec} rule for the $\SIwrC{x}{1}$ operation.
Again, we must open the invariant to gain $\color{darkblue}\RU\mapstoMem{x}{\Vals'}$, 
for a fresh set of values $\Vals'$ for which 
$\color{darkblue} \Vals' = \emptyset \lor \Vals' = \{1\}$ holds. Now, $\color{darkblue}\RU\mapstoMem{x}{\Vals'}$, 
together with the local points-to resource $\color{darkblue}\RU\mapstoCache{x}{\Cst}{\None}$ we obtained using the \ruleref{ru-start-spec} rule, is 
provided to the \ruleref{ru-write-spec} rule to obtain 
$\color{darkblue}\RU\mapstoMem{x}{\Vals' \cup \{1\}} \ast \RU\mapstoCache{x}{\Cst}{\Some{1}}$.
Notice that no matter which of the cases is true in the disjunction 
$\color{darkblue} \Vals' = \emptyset \lor \Vals' = \{1\}$,
we have that $\color{darkblue}\Vals' \cup \{1\}$ equals $\color{darkblue}\{1\}$ 
which is sufficient for restoring the client invariant 
again. We can hold on to $\color{darkblue}\RU\mapstoCache{x}{\Cst}{\None}$, but we will no longer need it at this point. 
The last operation is $\SIloop$, which is a non-terminating operation and which therefore trivially satisfies any specification (as we are using a partial-correctness logic; technically, in Iris this is proved using so-called L{\"o}b induction).

\paragraph{Read-transaction} As for the write-transaction, 
we initialize using the \ruleref{ru-init-client-spec} rule,
and thus we obtain ownership of the resource 
$\color{darkblue}\RU\CanStart{\Cst}$. 
Also, as for the write-transaction, reasoning about the start operation goes the same way 
by opening and closing the client invariant leaving us to continue with the resources 
$\color{darkblue} \RU\Active{\Cst}{\{x\}} \ast \RU\mapstoCache{x}{\Cst}{\None}$.
For the read operation $\SIrdC{x}$, we must open the client invariant to gain 
$\color{darkblue}\RU\mapstoMem{x}{\Vals}$ for a fresh set of values $\color{darkblue}\Vals$ for which 
$\color{darkblue} V = \emptyset \lor V = \{1\}$ holds. According to the postcondition of 
\ruleref{ru-read-spec}, as our local points-to resource points to $\color{darkblue}\None$, 
we get the following information about the return value $\color{darkblue}ow$ of the write operation: 
$\color{darkblue}(\exists v, ow = \Some{v} \land v \in \Vals) \lor ow = \None$. Having obtained this information 
we can close the invariant with $\color{darkblue}\RU\mapstoMem{x}{\Vals}$ and proceed to 
the assertion $\SIassert{v_x = \None \lor v_x = \Some{1}}$. As $v_x$ is equal to the return 
value of the write operation $\color{darkblue}ow$, we use the information 
$\color{darkblue}(\exists v, ow = \Some{v} \land v \in \Vals) \lor ow = \None$ together with our knowledge 
about the set $\color{darkblue}\Vals$, i.e., $\color{darkblue} V = \emptyset \lor V = \{1\}$, 
to conclude that the assertion does indeed hold.
Having concluded that the assertion holds, we are only left with reasoning about $\SIcommit$.
In the precondition of \ruleref{ru-commit-spec} we collect the connection state, 
$\color{darkblue} \RU\Active{\Cst}{\{x\}}$, 
the local points-to resource, $\color{darkblue}\RU\mapstoCache{x}{\Cst}{\None}$, and 
the global points-to resource from the invariant for a fresh $\color{darkblue} \Vals'$, 
$\color{darkblue}\RU\mapstoMem{x}{\Vals'}$. 
The postcondition gives us back $\color{darkblue}\RU\mapstoMem{x}{\Vals'}$ to close the invariant and 
the ability to start a new transaction in the form of the connection state 
$\color{darkblue}\RU\CanStart{\Cst}$.

\paragraph{Remarks}
Notice how the proof of the read uncommitted data example is modular: We use the node-local reasoning of Aneris, together with the client invariant, to reason about each transaction in isolation.
The proof is also modular in the sense that it only relies on the specification of read uncommitted, not the implementation thereof.
We further remark that it follows formally by the implication proofs in \Cref{sec:4,sec:6} that 
the example is also provable using read committed and snapshot isolation.


\section{Specifying Read Committed}
\label{sec:3}

In this section, we present our modular separation logic specification for read committed.
We also present the \emph{dirty read} example and the \emph{commit order} example
which we have proven using the specification.
Contrary to read uncommitted, read committed is of high practical value as it is the default isolation level
of leading database systems such as Microsoft SQL, Oracle or PostgreSQL.
A reader unfamiliar with transactional guarantees may be surprised by the wide spread of read committed
as, albeit being stronger than read uncommitted, read committed does not impose any order on the data being read.
Instead, read committed builds upon read uncommitted by adding the requirement that 
transactions are only allowed to read committed data.
Many formalizations specifying read committed, and other isolation levels for that sake, are 
not concerned with guarantees about aborting transaction. For example, \citet{Adya99,Crooks17} 
do not require that uncommitted transactions must read committed data. 
Allowing aborting transactions not to read committed data would significantly increase the complexity of our separation logic 
specification, and hence we have made the design choice that aborting transactions should also read 
committed data. In practice, imposing the requirement that uncommitted transactions may read \emph{any} uncommitted data 
is a weak requirement that will not exclude many, if any, implementations from realizing the specification. 
We remark that our specification still specifies read committed, 
it is just not the weakest specification there exists (an algorithm producing 
a subset of the valid executions under an isolation level does naturally still adhere 
to that isolation level).
We remark that an equivalent choice is made in \citet{vMVCC} when specifying serializability.
\begin{figure}[htbp]
  \centering
  \small
  \color{darkred}
  \begin{mathpar}
    \axiomH{rc-init-client-spec}
    { 
      \anhoare
        {
          \TRUE
        }
        {
          \SIinclC{\srvsa}
        } 
        { \begin{array}[t]{@{}l@{}}
          \Ret \Cst. \RC\CanStart{\Cst}
          \end{array}
          }
        {}{}
    }
    \and
    \axiomH{rc-init-kvs-spec}
    { 
      \anhoare
        { \begin{array}[t]{@{}l@{}}
          \RC\KVSinit
          \end{array}
        }
        { 
          \SIinsrv
        } 
        { \begin{array}[t]{@{}l@{}}
          \Ret \TT. \TRUE
        \end{array}}
        {\srvsa}{}
    }
    \and
    \axiomH{rc-start-spec}
    { 
      \anhoareVatomic
        { \begin{array}[t]{@{}l@{}}
          \RC\CanStart{\Cst} \ast
          \Sep_{(\Key,\, \Vals) \in\, \map} \RC\mapstoMem{\Key}{\Vals} 
        \end{array}
        }
        { \begin{array}[t]{@{}l@{}}
          \SIstartC{\Cst}
        \end{array}
        } 
        { \begin{array}[t]{@{}l@{}}
          \Ret \TT. \RC\Active{\Cst}{\dom(\map)} \asts \\
          \Sep_{(\Key,\, \Vals) \in\, \map} \Big(\RC\mapstoMem{\Key}{\Vals} \asts
          \RC\mapstoCache{\Key}{\Cst}{\None}\Big)
        \end{array}}
        {}{}
    }
    \and
    \axiomH{rc-read-spec}
    { 
      \anhoareVatomic
        { \begin{array}[t]{@{}l@{}}
          \RC\mapstoCache{\Key}{\Cst}{\ValueOption} \ast \RC\mapstoMem{\Key}{\Vals}
          \end{array}
        }
        { \begin{array}[t]{@{}l@{}}
          \SIrdC{\Cst}{\Key}
          \end{array}
        } 
        { \begin{array}[t]{@{}l@{}}
          \Ret \mathit{ow}. 
          \RC\mapstoCache{\Key}{\Cst}{\ValueOption} \ast \RC\mapstoMem{\Key}{\Vals} \asts \\
          \Big(\big(\ValueOption = \None ~ \land \\
          \quad ((\exists v, \mathit{ow} = \Some{v} \land v \in \Vals) ~ \lor \\ 
          \qquad \mathit{ow} = \None)\big) ~ \lor \\ 
          \big(\ValueOption \neq \None \land \mathit{ow} = \ValueOption \big)\Big)
        \end{array}}
        {}{}
    }
    \and
    \axiomH{rc-commit-spec}
    { 
      \anhoareVatomic
        { \begin{array}[t]{@{}l@{}}
          \RC\Active{\Cst}{\SetVal} \ast
          \SetVal = \dom ~ \map =  \dom ~ \Cache \asts \\[0.4em]
          \Sep_{(\Key,\, \Vals) \in\, \map} \RC\mapstoMem{\Key}{\Vals} ~ \asts
          \Sep_{(\Key,\, \ValueOption) \in\, \Cache} \RC\mapstoCache{\Key}{\Cst}{\ValueOption}
          \end{array}
        }
        { \begin{array}[t]{@{}l@{}}
          \SIcommitC{\Cst}
        \end{array}
        }
        { \begin{array}[t]{@{}l@{}}
          \Ret \val. \CanStart{\Cst} ~ \ast ~ \\[0.4em]
          \Big(\val = \TRUE \ast 
          \Sep_{\substack{(\Key,\, \Vals) \in\, \map \\ (\Key,\,\ValueOption) \in\, \Cache}}
          \RC\mapstoMem{\Key}{\CommitVals{\ValueOption}{\Vals}}\Big) ~ \lor ~ \\[0.7cm]
          \Big(\val = \FALSE \asts
          \Sep_{(\Key,\, \Vals) \in\, \map} \RC\mapstoMem{\Key}{\Vals}\big)
        \end{array}
        }
        {}{}
    }
    \and
    \axiomH{rc-write-spec}
    { 
      \anhoareVatomic
        {
         \begin{array}[t]{@{}l@{}}
          \RC\mapstoCache{\Key}{\Cst}{\ValueOption}
        \end{array}
        }
        {
          \SIwrC{\Cst}{\Key}{\val}
        } 
        { \begin{array}[t]{@{}l@{}}
          \Ret \TT. \RC\mapstoCache{\Key}{\Cst}{\Some \val}
          \end{array}
          }
        {}{}
    }
    \and
    \end{mathpar}
    \caption{Specification for read committed ($\color{darkred}\mathsf{RCSpecs}$).}
    \label{fig:specs_rc}
    \vspace{-0.4cm}
\end{figure}
\paragraph{Specification} The specification (\Cref{fig:specs_rc}) is based upon the same structure as the specification for read committed: 
local and global points-to resources for sets of values with the only
differences occurring in the \ruleref{rc-write-spec} and  the \ruleref{rc-commit-spec} rules. 
Hence, we present the read committed specification by 
highlighting the difference from the read uncommitted
specification in the \ruleref{rc-write-spec} and the \ruleref{rc-commit-spec} rules.

In comparison with \ruleref{ru-write-spec}, 
\ruleref{rc-write-spec} does not update the global points-to resource, and the update is simply recorded in the 
local points-to resource. This captures that a concurrently active transaction can not read the change until 
it is committed. In the \ruleref{rc-commit-spec}, if the commit is successful, the 
updates recorded in the local points-to resources are added to the global points-to resources 
using the $\color{darkred}\CommitValsSymb$ function, see \eqref{eq:update_vals}. In case the 
commit is unsuccessful, the key-value store state, i.e., the global points-to resources, remain 
unchanged and the changes of the transaction are lost.
\vspace{-0.1cm}
\begin{equation}
  \CommitVals{\ValueOption}{\Vals} \eqdef \textlog{match} ~ \ValueOption ~ \textlog{with} ~ \Some{v} => \Vals \cup v ~ | ~ \None => \Vals
\label{eq:update_vals}
\end{equation}
As for read uncommitted, observations can be made using $\SeenSymb$ resources. 
In fact, the exclusion property 
\eqref{eq:ru-exclusion-rule} and the creation property \eqref{eq:ru-creation-rule} 
holds for read committed too
(all resources should be prefixed with $\color{darkred}\RC$ instead of $\color{darkblue}\RU$).
To conclude the presentation of the read committed specification, we state the complete specification, 
similar to \eqref{eq:ru-setup_specs}, which formally defines the read committed specification ($\color{darkred}\mathsf{RCSpec}$), see \eqref{eq:rc-setup_specs}.
\begin{equation}
  \small
  \color{darkred}
  \mathsf{RCSpec} \eqdef 
  \pvs[][] ~
  \RC\KVSinit \ast
  \Sep_{\Key\, \in\, \Keys} \RC\mapstoMem{\Key}{\emptyset} \ast
  \mathsf{RCSpecs} \ast
  \RC\GlobalInv
  \color{black}
\label{eq:rc-setup_specs}
\end{equation}
\vspace{-0.4cm}

\label{sec:3.1}
\paragraph{Read Committed Example.} We have used the read committed specification to formally prove 
the \emph{dirty read} example \Cref{fig:dirty-read-example} and the \emph{commit order} example 
\Cref{fig:commit-order-example}.

\begin{figure}[h]
    \small
    \centering
    \begin{minipage}[b]{0.5\textwidth}
      \begin{align*}
        \left.
        \begin{aligned}
          & \SIstart\\
          & \SIwrC{x}{1}\\
          & \SIloop\\
          &
        \end{aligned}
        ~\middle\Vert~
        {
        \begin{aligned}
          & \SIstart\\
          & \SIletin{v_x}{\SIrdC{x}}\\
          & \SIassert{v_x = \None}\\
          & \SIcommit
        \end{aligned}
        }
        \right.
        \end{align*}
        \begin{align*}
          Inv &\eqdef{} \color{darkred} \exists V,~ \RC\mapstoMem{x}{V} ~ \ast V = \emptyset
        \end{align*}
        \vspace{-0.6cm}
    \caption{Dirty read example.}
    \label{fig:dirty-read-example}
    \vspace{-0.3cm}
  \end{minipage}
\end{figure}

The dirty read example builds upon the read uncommitted data example 
from \Cref{fig:read-uncommitted-example}, and is based upon 
the dirty read phenomenon in the literature 
(\citet{orig-SI,Adya99} and 1992 ANSI SQL standard), by removing the clause in the conjunction 
of the assertion that corresponds to the reading transaction seeing the uncommitted 
data from the writing transaction. This is as expected for 
read committed, since read committed disallows transactions to read uncommitted data.
Recall that a phenomenon is an example of an execution between concurrently executing transactions 
that should be prohibited;
in our example this is captured by the assertion  
showing that it is impossible.
The proof of the example is simple and uses the invariant shown in \Cref{fig:dirty-read-example}
which state that no value will be written to the $x$ variable.

\begin{figure}[h]
  \vspace{-0.25cm}
  \small
  \centering
  \begin{minipage}[b]{1\textwidth}
    \begin{align*}
    \left.
    \begin{aligned}
      & \SIstart\\
      & \SIwrC{x}{1}\\
      & \SIletin{v_y}{\SIrdC{y}}\\
      & \SIifthen{(v_y = \Some{1})}\\ 
      & \{\SIwrC{a}{1}\}\\
      & \SIcommit
    \end{aligned}
    ~\middle\Vert~
    {
    \left.
    \begin{aligned}
      & \SIstart\\
      & \SIwrC{y}{1}\\
      & \SIletin{v_x}{\SIrdC{x}}\\
      & \SIifthen{(v_x = \Some{1})}\\
      & \{\SIwrC{b}{1}\}\\
      & \SIcommit
    \end{aligned}
    ~\middle\Vert~
    \begin{aligned}
      & \SIstart\\
      & \SIletin{v_a}{\SIrdC{a}}\\
      & \SIletin{v_b}{\SIrdC{b}}\\
      & \SIassert{!(v_a = \Some{1} \land v_b = \Some{1})}\\
      & \SIcommit \\
      &
    \end{aligned}
    \right.
    }
    \right.
    \end{align*}
    \caption{Commit order example.}
    \label{fig:commit-order-example}
    \vspace{-0.3cm}
  \end{minipage}
\end{figure}

The commit order example asserts the absence of a situation 
in which two transactions both are seeing data committed by the other transaction. 
This should not be allowed to occur as there must be an order in which transactions commit.
The first two transactions in the dirty read example 
revolve around the same structure: First the value 1 is written into a key 
that the other transaction reads from and checks whether it was written to using an 
if-statement. 
If this was the case, the transactions write to another key (respectively $a$ and $b$)
to signify that they saw the write of the other transaction. The last transaction 
reads from $a$ and $b$ and asserts that the transactions could not possible both 
have seen each others writes (the case where both $a$ and $b$ has been written to).

We remark that it follows from the implication proof in \Cref{sec:6} that 
the two examples in this section are also provable for snapshot isolation.

\section{Read Committed Implies Read Uncommitted}
\label{sec:4}

Given our formal separation logic specifications of read uncommitted and read committed, we can
now formally prove that read committed implies read uncommitted:
\begin{theorem}[Read committed implies read uncommitted]
  \label{rc_ru_theorem}
  The specification for read committed \eqref{eq:rc-setup_specs} implies the specification for read uncommitted \eqref{eq:ru-setup_specs}: 
  \[\color{darkred}\mathsf{RCSpec} \color{black}\rightarrow \color{darkblue}\mathsf{RUSpec}.\] 
\vspace{-0.4cm}
\end{theorem}
\paragraph*{Proof Structure} The proof, consisting of approximately 800 lines of Rocq proof code,
centers around closing the gap between read uncommitted and read committed that comes from the fact 
that in read uncommitted the global state is updated immediately at the time of writing values, whereas for read
committed, the global state is not updated until the transaction commits successfully.
The proof proceeds naturally by first assuming the read committed specification 
$\color{darkred}\mathsf{RCSpec} \color{black}$ \eqref{eq:rc-setup_specs}, 
which consists of the global invariant $\color{darkred}\RC\GlobalInv\color{black}$, 
fresh global points-to resources
$\color{darkred}\Sep_{(\Key,\, \Vals) \in\, \Keys} \RC\mapstoMem{\Key}{\emptyset}\color{black}$, 
an initialization resource $\color{darkred}\RC\KVSinit\color{black}$ for the key-value store
and the read committed per operation specifications $\color{darkred}\mathsf{RCSpecs} \color{black}$ 
(\Cref{fig:specs_rc}). To prove $\color{darkblue}\mathsf{RUSpec}\color{black}$,
we must naturally show the equivalent resources and specifications for read uncommitted 
(including the exclusion \eqref{eq:ru-exclusion-rule} and creation \eqref{eq:ru-creation-rule} properties).
Now, we need to define all the resources used in 
$\color{darkblue}\mathsf{RUSpec}\color{black}$. In doing this, 
we will have all the resources found in $\color{darkred}\mathsf{RCSpec} \color{black}$ 
at our disposal. 
We define the key-value store resource for initialization 
to be the same as the corresponding read committed resource, 
but for the global invariant and the global and local points-to resources
we use a more involved definition with \emph{resource algebras}.
Resource algebras are the building blocks of Iris,
see \citet{iris2} for details.
All we need to know here is that using resource algebras, we can create custom predicates
 and rules to relate these predicates 
 (usually we will refer to these custom predicates as resources algebras).
Having defined all resources, the largest part of the proof revolves around showing the 
per operation specifications of read uncommitted, i.e., $\color{darkblue}\mathsf{RUSpecs} \color{black}$.
When proving each specification, we get to use
the corresponding specification for read committed from 
$\color{darkred}\mathsf{RCSpecs} \color{black}$ as an assumption. 
This is helpful, because
we first assume the read uncommitted resources in the precondition of the specification
and these we have defined in terms
of the read committed resources and resource algebras. 
We thus use the read committed specification for the particular operation to transform the read committed 
resources, while the resource algebras are transformed by hand, to meet the post 
condition of the read uncommitted specification. 

We remark that while the proof follows a quite natural path, 
it is non-trivial to come up with a definition 
of the read uncommitted resources in terms of the read committed resources that makes the proof work.
We refer to the Rocq formalization for the details of the whole proof 
and continue here with an excerpt regarding the write specification and insights about how the proof 
goes at commit time for the commit specification.

\paragraph*{Proof Excerpt} Our goal is to prove the \ruleref{ru-write-spec} rule which uses 
the resources $\color{darkblue} \RU\mapstoMem{x}{V}$ and 
$\color{darkblue}\RU\mapstoCache{\Key}{\Cst}{\ValueOption}$. 
We get to assume the \ruleref{rc-write-spec} rule and can define 
the read uncommitted resources, which we will refer to as the \emph{wrapped} resources,
using the equivalent resources for read committed, 
i.e., $\color{darkred} \RC\mapstoMem{x}{V}$ and 
$\color{darkred}\RC\mapstoCache{\Key}{\Cst}{\ValueOption}$.

To define the wrapped resources, we use a \emph{ghost theory}, which is 
a set of separation logic predicates defined using resource algebras, 
together with rules that relate the different predicates. 
Our ghost theory consists of two predicates, $\AuthSet{V}$ and $\FragSet{V}$, on sets of values $V$.
Intuitively, the \emph{authorative} part $\AuthSet{V}$ can be initialized with the empty set, 
i.e., $V = \emptyset$, and will at all times hold all the values that have been added since $V$ was the empty set. 
This is seen in \eqref{eq:set_addition} where, given the global invariant and $\AuthSet{V}$, one can make an update 
(signified by $\pvs[]{}$) by adding an element to $V$ resulting in $\AuthSet{V \cup \val}$. 
We remark that unless explicitly stated otherwise, resources from an invariant in Iris 
can be used to prove a proposition below the update modality
as long as the invariant still holds after the update.
\begin{equation}
  \small
  \color{darkblue} \RU\GlobalInv \color{black} \ast \AuthSet{V} 
  \proves \pvs[]{} ~ \color{darkblue} \RU\GlobalInv \color{black} \ast \AuthSet{V \cup \val} \ast \FragSet{V \cup \val}
  \color{black}
 \label{eq:set_addition}
\end{equation}
At the same time, one gains the \emph{fragmental} part $\FragSet{V \cup \val}$. As such, 
$\FragSet{V}$, for any $V$, represents the values of $\AuthSet{V}$ at some point in the past
(note that $\eqref{eq:set_addition}$ does not update old fragmental parts when new values are added, and values can not be removed).
Naturally, this means that at any given point, we can conclude that the set of values in 
a fragmental resource is a subset of the values in an authoritative resource, see \eqref{eq:set_inclusion}.
\begin{equation}
  \small
  \color{darkblue} \RU\GlobalInv \color{black} \ast \AuthSet{V} \ast \FragSet{V'}
  \proves \color{darkblue} \RU\GlobalInv \color{black} \ast \AuthSet{V} \ast \FragSet{V'} \ast V' \subseteq V
  \color{black}
 \label{eq:set_inclusion}
\end{equation}
Having defined our ghost theory, we can define the wrapped resources used in the write specification
 \Cref{fig:wrapped-resource}.
\begin{figure}[h]
  \small
  \centering
    \begin{align*}
      \begin{aligned}
        \color{darkblue} \RU\mapstoMem{x}{V} \color{black} \eqdef & ~
        \exists V', V' \subseteq V \ast \color{darkred} \RC\mapstoMem{x}{V'} \color{black} \ast \AuthSet{V} \\
        \color{darkblue} \RU\mapstoCache{\Key}{\Cst}{\ValueOption} \color{black} \eqdef & ~ \exists V, 
        ((\exists \val, \ValueOption = \Some{\val} \land \val \in V) \lor \ValueOption = \None) 
        \ast \color{darkred} \RC\mapstoCache{\Key}{\Cst}{\ValueOption} \color{black} \ast \FragSet{V}
      \end{aligned}
      \end{align*}
  \vspace{-0.2cm}
  \caption{Wrapped resources.}
  \label{fig:wrapped-resource}
\end{figure}
Recall, $\color{darkblue} \RU\mapstoCache{\Key}{\Cst}{\ValueOption} \color{black}$ and 
$\color{darkred} \RC\mapstoCache{\Key}{\Cst}{\ValueOption} \color{black}$ are used in the same way in 
read uncommitted and read committed: 
$\color{darkblue}\ValueOption\color{black}/\color{darkred}\ValueOption\color{black}$ is 
the latest value written (if a write has been made) by the current active transaction,
and it is updated using the write specification. In contrast, 
$\color{darkblue} \RU\mapstoMem{x}{V} \color{black}$ and 
$\color{darkred} \RC\mapstoMem{x}{V} \color{black}$, holding the global state, are not used in the same way: 
For read uncommitted the global state 
is updated immediately at the time of writing the value, 
and for read committed the global state is not updated until the transaction commits successfully. 
Therefore, the wrapped resources for read uncommitted in \Cref{fig:wrapped-resource} 
are constructed to hide the fact that values are not propagated to the global state until commit time. 
To see this, observe that inside the definition of 
$\color{darkblue} \RU\mapstoMem{x}{V} \color{black}$ lies 
$\color{darkred} \RC\mapstoMem{x}{V'} \color{black}$ with $V'$ being a subset of $V$. 
The set $V$ contains the committed and uncommitted values whereas $V'$ only contains the committed values.  
$\AuthSet{V}$ is then used to keep track of all the uncommitted values. 
Further, $\color{darkblue} \RU\mapstoCache{\Key}{\Cst}{\ValueOption} \color{black}$ is defined using the equivalent 
resource $\color{darkred} \RC\mapstoCache{\Key}{\Cst}{\ValueOption} \color{black}$ 
where it is also stated that if $\ValueOption = \Some{\val}$ for some value $\val$, then 
$\val$ must be in the set of all values (captured using $\FragSet{V}$ and the requirement $\val \in V$ for some $V$).
In the proof of the write specification, which we will sketch next,
it is not evident that we
the need our ghost theory since we only update it, but we do not use it to draw any 
conclusions. The necessity of the ghost theory appears at commit time: 
As $V'$ in $\color{darkred} \RC\mapstoMem{x}{V'} \color{black}$ is updated with $\val$
from $\color{darkred} \RC\mapstoCache{\Key}{\Cst}{\Some{\val}} \color{black}$
at commit time for read committed, 
and $\color{darkblue} \RU\mapstoMem{x}{V} \color{black}$ remains the same at commit time for read uncommitted,
we use the fact that there exists some $V''$ with $\val \in V''$, 
for which $\FragSet{V''}$ holds, per the definition of our wrapped resource, 
which implies $\val$ is also in $V$ using rule \eqref{eq:set_inclusion} of our ghost theory 
(the resource $\AuthSet{V}$ comes from inside the definition of $\color{darkblue} \RU\mapstoMem{x}{V} \color{black}$). 
Thus, at commit time, even though $\color{darkred} \RC\mapstoMem{x}{V'} \color{black}$ 
changes inside the wrapped resource $\color{darkblue} \RU\mapstoMem{x}{V} \color{black}$, 
we can conclude that the added value is already in $V$, and $V$ can remain unchanged,
 due to the ghost theory construction, 
which is needed to prove $\ruleref{ru-commit-spec}$.
In this sense, the wrapped resource are constructed to hide the fact 
that values are not propagated until commit time.

The write specification we end up having to prove (when unwrapping the wrapped resources) is shown in 
\Cref{fig:wrapped-write-spec}. 
\begin{figure}[ht]
  \small
  \centering
  \begin{mathpar}
    \axiomH{}
    { 
      \anhoareVatomic
        {
         \begin{array}[t]{@{}l@{}}
          (\exists V, 
          ((\exists \val', \ValueOption = \Some{\val'} \land \val' \in V) \lor \ValueOption = \None) 
          \ast \color{darkred} \RC\mapstoCache{\Key}{\Cst}{\ValueOption} \color{black} \ast \FragSet{V}) \asts \\ 
          (\exists V', V' \subseteq V \ast \color{darkred} \RC\mapstoMem{x}{V'} \color{black} \ast \AuthSet{V})
        \end{array}
        }
        {
          \SIwrC{\Cst}{\Key}{\val}
        } 
        { \begin{array}[t]{@{}l@{}}
          \Ret \TT. 
          (\exists V, 
          ((\exists \val', \Some{\val} = \Some{\val'} \land \val' \in V) \lor \Some{\val} = \None) 
          \ast \color{darkred} \RC\mapstoCache{\Key}{\Cst}{\Some{\val}} \color{black} \ast \FragSet{V}) \asts \\ 
          (\exists V', V' \subseteq V \cup \val \ast \color{darkred} 
          \RC\mapstoMem{x}{V'} \color{black} \ast \AuthSet{V \cup \val})
          \end{array}
          }
        {}{}
    }
  \end{mathpar}
  \caption{Unwrapped write specification.}
  \label{fig:wrapped-write-spec}
\end{figure}

\noindent In the precondition, when instantiating the existential quantifiers, 
the relevant resources we get to assume are
$\color{darkred} \RC\mapstoCache{\Key}{\Cst}{\ValueOption} \color{black} \ast
 \color{darkred} \RC\mapstoMem{x}{V'} \color{black} \ast \AuthSet{V}$ with $V' \subseteq V$. 
Using the read committed write specification \ruleref{rc-write-spec}, we can transform the resource 
$\color{darkred} \RC\mapstoCache{\Key}{\Cst}{\ValueOption}$ into
$\color{darkred} \RC\mapstoCache{\Key}{\Cst}{\Some{\val}} \color{black}$. 
Using \eqref{eq:set_addition}, we can transform $\AuthSet{V}$ into 
$\AuthSet{V \cup \val} \ast \FragSet{V \cup \val}$. 
Hence, we can now satisfy the postcondition of \Cref{fig:wrapped-write-spec}, using 
$\color{darkred} \RC\mapstoCache{\Key}{\Cst}{\Some{\val}} \color{black} \ast 
\color{darkred} \RC\mapstoMem{x}{V'} \color{black}
\ast \AuthSet{V \cup \val} \ast \FragSet{V \cup \val}$, which concludes the proof sketch of the 
write specification.


\section{Specifying Snapshot Isolation}
\label{sec:5}

In this section, we first present our modular separation logic specification for snapshot isolation, and then 
we present three examples based on phenomena from the literature, \emph{write skew}, \emph{read skew} and \emph{non-repeatable read}, 
in addition to a bank transfer example which we have all proven using the specification.
Snapshot isolation, like read committed, is an isolation level with practical value, but it comes with
much stronger guarantees about what transactions are allowed to read.

Snapshot isolation, as presented in \citet{orig-SI}, makes use of snapshots. 
A snapshot is defined as the state of the database at a given time.
Snapshot isolation makes a distinction between the start snapshot, 
on which a transaction reads, and the commit snapshot, 
on which the transaction attempts to commit its updates.
All commits happen in a globally defined linear order.
The start and commit snapshots are taken at the time of the 
start operation and the commit operation respectively.
In this paper, the snapshots will be the most recent snapshots at the time they are a captured.
This can be referred to as "strong snapshot isolation" \citep{strong-SI}, 
as one can relax the criteria on the start snapshot to be any valid snapshot 
at the time of the start operation --- not necessarily the most recent.
\footnote{In \citet{orig-SI} starting snapshots are allowed to be any previous snapshot and 
not necessarily the most recent. 
Specifications for the weaker variant would be much harder to work with, 
as you would get a disjunction of all possible snapshots to reason about when starting a transaction.}
Naturally, the start snapshot and the commit snapshot can be different as other concurrent transactions can commit in the time between the snapshots.
This leads us to the crucial commit-check of snapshot isolation, which expresses when a transaction can commit: \\
\begin{quote}
  \vspace{-0.3cm}
  \emph{(Snapshot Isolation Commit-check)} A transaction is allowed to commit only if it has no write conflicts with any transaction
  committed between its own start snapshot and its commit snapshot. \\
  \vspace{-0.3cm}
\end{quote}
\noindent As for the previous isolation levels, our separation logic specification 
will be based around local and global points-to resources, 
but this time the global points-to resources do not point to a set of values but \emph{histories}. 
Histories are lists of values with the last element representing the latest committed value to the key. 
Histories, as opposed to sets, are needed to express the snapshot isolation commit check: we can use
the histories from a start snapshot and a commit snapshot to
conclude whether writes have happened between the two snapshots. 

The initialization specifications, \ruleref{si-init-client-spec} and \ruleref{si-init-kvs-spec}, 
are identical to the ones of the previous isolation levels. We now
describe the specifications of the start, read, write and commit operations for 
snapshot isolation which can all be found in \Cref{fig:specs_si}. 
\begin{figure}[htbp]
  \centering
  \small
  \color{darkgreen}
  \begin{mathpar}
    \axiomH{si-init-client-spec}
    { 
      \anhoare
        {
          \TRUE
        }
        {
          \SIinclC{\srvsa}
        } 
        { \begin{array}[t]{@{}l@{}}
          \Ret \Cst. \SI\CanStart{\Cst}
          \end{array}
          }
        {}{}
    }
    \and
    \axiomH{si-init-kvs-spec}
    { 
      \anhoare
        { \begin{array}[t]{@{}l@{}}
          \SI\KVSinit
          \end{array}
        }
        { 
          \SIinsrv
        } 
        { \begin{array}[t]{@{}l@{}}
          \Ret \TT. \TRUE
        \end{array}}
        {\srvsa}{}
    }
    \and
    \axiomH{si-start-spec}
    { 
      \anhoareVatomic
        { \begin{array}[t]{@{}l@{}}
          \SI\CanStart{\Cst} \ast
          \Sep_{(\Key,\, \Hist) \in\, \map} \SI\mapstoMem{\Key}{\Hist}
        \end{array}
        }
        { \begin{array}[t]{@{}l@{}}
          \SIstartC{\Cst}
        \end{array}
        } 
        { \begin{array}[t]{@{}l@{}}
          \Ret \TT. \SI\Active{\Cst}{\map} \asts\\
          \Sep_{(\Key,\, \Hist) \in\, \map} \Big(\SI\mapstoMem{\Key}{\Hist} \asts
          \SI\mapstoCache{\Key}{\Cst}{\HistVal{\Hist}} \asts \\
          \hspace{1.4cm} \SI\KeyUpdStatus{\Cst}{\Key}{\FALSE} \Big)
        \end{array}}
        {}{}
    }
    \and
    \axiomH{si-write-spec}
    { 
      \anhoareVatomic
        {
         \begin{array}[t]{@{}l@{}}
          \SI\mapstoCache{\Key}{\Cst}{\ValueOption} \asts \\
          \SI\KeyUpdStatus{\Cst}{\Key}{\Boolean}
        \end{array}
        }
        {
          \SIwrC{\Cst}{\Key}{\val}
        } 
        { \begin{array}[t]{@{}l@{}}
          \Ret \TT. \SI\mapstoCache{\Key}{\Cst}{\Some \val} \asts \\
                                      \SI\KeyUpdStatus{\Cst}{\Key}{\TRUE}
          \end{array}
          }
        {}{}
    }
    \and
    \axiomH{si-commit-spec}
    { 
      \anhoareVatomic
        { \begin{array}[t]{@{}l@{}}
          \SI\Active{\Cst}{\Snapshot} \ast \dom ~ \map = \dom ~ \Snapshot =  \dom ~ \Cache \asts \\[0.4em]
          \Sep_{(\Key,\, \Hist) \in\, \map} \SI\mapstoMem{\Key}{\Hist} ~ \asts
          \Sep_{(\Key,\, (\ValueOption,\, \Boolean)) \in\, \Cache} 
            \big(\SI\mapstoCache{\Key}{\Cst}{\ValueOption} \ast \SI\KeyUpdStatus{\Cst}{\Key}{\Boolean}\big)
          \end{array}
        }
        { \begin{array}[t]{@{}l@{}}
          \SIcommitC{\Cst}
        \end{array}
        }
        { \begin{array}[t]{@{}l@{}}
          \Ret \val. \SI\CanStart{\Cst} ~ \ast ~ \\[0.4em]
          \Big(\val = \TRUE \ast \SI\CanCommit{\map}{\Snapshot}{\Cache} \ast 
          \Sep_{\substack{(\Key,\, \Hist) \in\, \map \\ 
          (\Key,\, p) \in\, \Cache}} \SI\mapstoMem{\Key}{\CommitHist{p}{\Hist}}\Big) ~ \lor ~ \\[0.7cm]
          \Big(\val = \FALSE \ast \lnot \SI\CanCommit{\map}{\Snapshot}{\Cache} ~ \asts
          \Sep_{(\Key,\, \Hist) \in\, \map} \SI\mapstoMem{\Key}{\Hist}\Big)
        \end{array}
        }
        {}{}
    }
    \and
    \axiomH{si-read-spec}
    { 
      \anhoareVatomic
        { \begin{array}[t]{@{}l@{}}
          \SI\mapstoCache{\Key}{\Cst}{\ValueOption}
          \end{array}
        }
        { \begin{array}[t]{@{}l@{}}
          \SIrdC{\Cst}{\Key}
          \end{array}
        } 
        { \begin{array}[t]{@{}l@{}}
          \Ret \ValueOption. 
          \SI\mapstoCache{\Key}{\Cst}{\ValueOption}
        \end{array}}
        {}{}
    }
    \and
    \end{mathpar}
    \caption{Specification for snapshot isolation ($\color{darkgreen}\mathsf{SISpecs}$).}
    \label{fig:specs_si}
    \vspace{-0.3cm}
\end{figure}

\paragraph{Start} 
In snapshot isolation, when starting a transaction, we take a snapshot of the database.
It is on the basis of this snapshot that the sequential reasoning inside the transactions happens: 
If the transaction does not include any writes, then all reads happen from this snapshot.
In the precondition of \ruleref{si-start-spec}, 
the client will provide its connection state together with the state of the key-value store
in form of global points-to resources,
which is typically obtained from an invariant.
The start operation then changes the connection state to active such that it holds 
information about its start snapshot $(\color{darkgreen}\SI\Active{\Cst}{\map}\color{black})$.
As the sequential reasoning inside transactions happens based on the start snapshot, 
all the local points-to resources are created with values equal to the current state of 
the key-value store.
As the start of a transaction does not apply any updates, the global points-to resources are 
given back without modifications for the client to reestablish its invariant.
When the transaction starts modifying the local points-to resources by doing writes, 
we lose the ability to distinguish the original snapshot from client changes.
That is why the \ruleref{si-start-spec} rule includes the resource 
$\color{darkgreen}\SI\KeyUpdStatus{\Cst}{\Key}{\FALSE}$ in the postcondition.
This resource states, for a particular key, whether the client did an update to this key.
When talking about the \ruleref{si-write-spec} rule next, we will see how the resource is updated and that it is 
this point that distinguishes our write specification from a store operation for the heap.

\paragraph{Read and Write} The read and write operations are not concerned 
with the global points-to resources, as they are working on a snapshot 
of the database, initially captured at start time in the local-points to resources. 
Hence, a local points-to resource $\color{darkgreen}\SI\mapstoCache{\Key}{\Cst}{\mathit{vo}}$ is used in 
the \ruleref{si-read-spec} rule to obtain a specification that is similar to the standard specification of a load operation on the heap.
Likewise, the \ruleref{si-write-spec} rule is manipulating the points-to predicate similarly to a store operation for the heap, 
except the change is recorded in the $\color{darkgreen}\SI\KeyUpdStatus{\Cst}{\Key}{\Boolean}$ resource which we need at commit time.

\paragraph{Commit}
Before delving into the details of the \ruleref{si-commit-spec} rule, we remind the reader 
how transactions are supposed to commit under snapshot isolation.
According to the snapshot isolation commit-check, we have to check if there are any 
write-conflicts between what the transaction wrote and what other concurrent transactions 
have done in the meantime.
Therefore, to use the \ruleref{si-commit-spec} rule, one must provide 
(1) the update status for each key in the starting snapshot, in the form of 
$\color{darkgreen}\SI\KeyUpdStatus{\Cst}{\Key}{\Boolean}$, together with the updated value 
from the local points-to resources ($\color{darkgreen}\Cache$ holds the information for these resources); 
(2) the starting snapshot of the transaction, in the form of 
$\color{darkgreen}\SI\Active{\Cst}{\Snapshot}$, where $\color{darkgreen}\Snapshot$ 
is exactly the snapshot gained from using the \ruleref{si-start-spec} rule; 
(3) the current snapshot of the key-value store, that is, 
the commit snapshot denoted as $\color{darkgreen}\map$, in the form of global points-to 
resources for all the keys in the start snapshot.
Having provided the necessary resources in the precondition of the \ruleref{si-commit-spec} rule, 
the postcondition expresses that the operation can proceed in one of two ways: 
Either the transaction unsuccessfully commits and the state of the key-value store remains the same 
as the current state, or the transaction successfully commits and the state of the 
key-value store is updated with the updates of the transaction.
To express the update to the key-value store we use the function $\color{darkgreen}\CommitHistSymb$, 
which takes as arguments both a pair consisting of value and a boolean, and a history.
The history is updated, by appending the value to it, if the update-status is true 
(the update status being the boolean argument).
In either case of the postcondition, evidence is provided as to why the case happened: 
If the transaction committed successfully, 
$\color{darkgreen}\SI\CanCommit{\map}{\Snapshot}{\Cache}$ will be true, 
which expresses that there are no write-conflicts with other transactions.
Dually, if the transaction is unsuccessful in committing, 
$\color{darkgreen}\SI\CanCommit{\map}{\Snapshot}{\Cache}$ is false, 
entailing at least one write-conflict with another transaction.
In \Cref{fig:can_commit}, we display the predicate for which $\color{darkgreen}\SI\CanCommitSymb$ 
is a decidable procedure.
The predicate captures that if the transaction did an update, 
($\MapLookup{\Cache}{\Key} = \SOME{(p, \TRUE)}$), 
then the start snapshot and the commit snapshot must be equal 
($\MapLookup{\map}{\Key} = \MapLookup{\Snapshot}{\Key}$).
\vspace{-0.2cm}
\begin{figure}[htbp]
  \small
  \centering
  \begin{mathpar}
    \CanCommitPredicate{\map}{\Snapshot}{\Cache} \eqdef
      \forall \Key \in \Keys, ~ \MapLookup{\Cache}{\Key} = \SOME{(p, \TRUE)}
      \Rightarrow \MapLookup{\map}{\Key} = \MapLookup{\Snapshot}{\Key}.
  \end{mathpar}
  \vspace{-0.6cm}
  \caption{The $\CanCommitPredicateSymb$ predicate for which the 
    $\CanCommitSymb$ function is a decidable procedure.} 
  \label{fig:can_commit}
\end{figure}
\vspace{-0.2cm}
\paragraph{Observations} This completes our explanation of our specifications of 
the snapshot isolation operations. As for read uncommitted and read committed, clients 
are able to make observations using the $\SeenSymb$ resource. In the exclusion property \eqref{eq:si-exclusion-rule} 
and the creation property \eqref{eq:si-creation-rule} below, 
we see how $\SeenSymb$ resources work on histories instead of sets of values.
Moreover, we see that from the exclusion property, a client is able to conclude that an observed history must 
be a prefix of the history representing the key-value store state. In \Cref{sec:app_util}, we 
discuss the differences between observations made in read uncommitted, read committed and snapshot isolation.
\begin{equation}
  \color{darkgreen}
  \small
    \RU\GlobalInv \ast \SI\Seen{\Key}{\Hist'} \ast \SI\mapstoMem{\Key}{\Hist} 
    \proves \pvs[]{} ~ \SI\mapstoMem{\Key}{\Hist} \ast \Hist' \leq \Hist
  \color{black}
    \label{eq:si-exclusion-rule}
\end{equation}
\vspace{-0.3cm}
\begin{equation}
  \color{darkgreen}
  \small
    \SI\GlobalInv \ast \SI\mapstoMem{\Key}{\Hist} 
    \proves \pvs[]{} ~ \SI\mapstoMem{\Key}{\Hist} \ast \SI\Seen{\Key}{\Hist}
  \color{black}
    \label{eq:si-creation-rule}
\end{equation}
\noindent To conclude the presentation of the snapshot isolation specification, we will a state the complete specification 
similar to \eqref{eq:ru-setup_specs} and \eqref{eq:rc-setup_specs} that formally defines the snapshot specification 
($\color{darkgreen}\mathsf{SISpec}$), see \eqref{eq:si-setup_specs}.
\begin{equation}
  \small
  \color{darkgreen}
  \mathsf{SISpec} \eqdef 
    \pvs[][] ~
    \SI\KVSinit \ast
    \Sep_{\Key\, \in\, \Keys} \SI\mapstoMem{\Key}{\HistEmpty} \ast
    \mathsf{SISpecs} \ast
    \SI\GlobalInv 
  \color{black}
\label{eq:si-setup_specs}
\end{equation}
\vspace{-0.3cm}

\label{sec:5.1}
\paragraph{Snapshot Isolation Examples.} For snapshot isolation, we have proven three examples based on phenomena from 
the literature (\citet{orig-SI,Adya99} and 1992 ANSI SQL standard) and a bank transfer example. 
The examples based on phenomena are the \emph{write skew} example in \Cref{fig:write-skew-example}, 
the \emph{read skew} example in \Cref{fig:read-skew-example} and 
the \emph{non-repeatable read} example in \Cref{fig:non-repeatable-read-example}.
The formal proofs of these three examples proceed in a manner similar to the proof shown
for read uncommitted.
For details see the accompanying Rocq formalization. 
In addition to the examples for snapshot isolation presented in this section, we have verified a number 
of additional examples, displaying important properties of snapshot isolation, which can be found in 
\Cref{sec:app_portfolio}.

\begin{figure}[h!]
  \small
  \centering
  \vspace{-0.3cm}
    \begin{minipage}[b]{0.29\textwidth}
    \begin{align*}
    \left.
    \begin{aligned}
      & \SIstart\\
      & \SIrdC{y}\\
      & \SIwrC{x}{1}\\
      & \SIassert{\SIcommit}
    \end{aligned}
    ~\middle\Vert~
    {
    \begin{aligned}
      & \SIstart\\
      & \SIrdC{x}\\
      & \SIwrC{y}{1}\\
      & \SIassert{\SIcommit}
    \end{aligned}
    }
    \right.
    \end{align*}
    \caption{Write skew.}
    \label{fig:write-skew-example}
  \end{minipage}\hspace*{-0.2em}%
  \begin{minipage}[b]{0.38\textwidth}
    \begin{align*}
      \left.
      \begin{aligned}
        & \SIstart\\
        & \SIwrC{x}{1}\\
        & \SIwrC{y}{1}\\
        & \SIassert{\SIcommit} \\
        &
      \end{aligned}
      ~\middle\Vert~
      {
      \begin{aligned}
        & \SIstart\\
        & \SIletin{v_x}{\SIrdC{x}}\\
        & \SIletin{v_y}{\SIrdC{y}}\\
        & \SIassert{v_x = v_y} \\
        & \SIassert{\SIcommit}
      \end{aligned}
      }
      \right.
      \end{align*}
    \caption{Read skew.}
    \label{fig:read-skew-example}
  \end{minipage}\hspace*{-0.2em}%
  \begin{minipage}[b]{.29\textwidth}
    \begin{align*}
      \left.
      \begin{aligned}
        & \SIstart\\
        & \SIwrC{x}{1}\\
        & \SIassert{\SIcommit} \\
        & \\
        &
      \end{aligned}
      ~\middle\Vert~
      {
      \begin{aligned}
        & \SIstart\\
        & \SIletin{v_1}{\SIrdC{x}}\\
        & \SIletin{v_2}{\SIrdC{x}}\\
        & \SIassert{v_1 = v_2}\\ 
        & \SIassert{\SIcommit}
      \end{aligned}
      }
      \right.
      \end{align*}
    \caption{Non-repeatable read.}
    \label{fig:non-repeatable-read-example}
  \end{minipage}
  \vspace{-0.5cm}
\end{figure}

The write skew example asserts that all executions between a transaction
that reads from one key and writes to another key 
and another transaction that does the same, but swaps the keys, will commit. 
Note that even the execution in which both transactions share the same snapshot,
in form of the initial snapshot, succeeds in snapshot isolation whereas it is
not allowed under serializability because serializability requires there to be an order amongst 
transactions. Naturally, proving the assertions in this example 
relies on the inclusion of the $\CanCommitPredicateSymb$ predicate in the \ruleref{si-commit-spec} rule.
The read skew example asserts that a transaction will either see all the writes or no writes 
from other transactions. Finally, the non-repeatable read example asserts that once a 
transaction sees a write from another transaction it can not unsee it.

\paragraph{Bank transfer} As opposed to the arguably contrived examples based on phenomena from the 
literature, the bank transfer example in \Cref{fig:bank-transfer-example} represents a common use case. 
Namely, the use case in which an amount of money is transferred from one account to another 
account in a banking system, given that there is sufficient funds in the source account.
The example in \Cref{fig:bank-transfer-example} contains a single transaction, 
which we imagine could be part of a larger system, and in comments it 
is shown how each line of code interacts with the separation logic resources, and, 
most importantly, how the invariant of the system is updated.
The invariant for this example, which could be a part

\begin{wrapfigure}{R}{0.5\textwidth}
  \small
  \centering
  \begin{minipage}[t]{0.5\textwidth}
    \begin{align*}
      \begin{aligned}
        & \color{darkgreen} // ~ \fbox{$\SI\mapstoMem{\mathit{src}}{\Hist_{\mathit{src}} \listapp [v_{\mathit{src}}]} \ast \SI\mapstoMem{\mathit{dst}}{\Hist_{\mathit{dst}} \listapp [v_{\mathit{dst}}]}$}\\
        & \SIstart\\
        & \color{darkgreen} // ~ \SI\mapstoCache{\mathit{src}}{\Cst}{v_{\mathit{src}}} \ast \SI\mapstoCache{\mathit{dst}}{\Cst}{v_{\mathit{dst}}}\\
        & \SIletin{bal_{\mathit{src}}}{\SIrdC{\mathit{src}}}\\
        & \color{darkgreen} // ~ \SI\mapstoCache{\mathit{src}}{\Cst}{v_{\mathit{src}}} \ast \SI\mapstoCache{\mathit{dst}}{\Cst}{v_{\mathit{dst}}} \land bal_{\mathit{src}} = v_{\mathit{src}}\\
        & \SIifthen{(bal_{\mathit{src}} \ge \mathit{amount})}\\ 
        & \{ \\
        & \quad \color{darkgreen} // ~ \SI\mapstoCache{\mathit{src}}{\Cst}{v_{\mathit{src}}} \ast \SI\mapstoCache{\mathit{dst}}{\Cst}{v_{\mathit{dst}}} ~ \land \\
        & \qquad \color{darkgreen} ~ bal_{\mathit{src}} = v_{\mathit{src}} \land v_{\mathit{src}} \ge \mathit{amount}\\
        & \quad \SIwrC{\mathit{src}}{(bal_{\mathit{src}} - \mathit{amount})} \\
        & \quad \color{darkgreen} // ~ \SI\mapstoCache{\mathit{src}}{\Cst}{v_{\mathit{src}} - \mathit{amount}} \ast \SI\mapstoCache{\mathit{dst}}{\Cst}{v_{\mathit{dst}}} ~ \land\\
        & \qquad \color{darkgreen} bal_{\mathit{src}} = v_{\mathit{src}} \land v_{\mathit{src}} \ge \mathit{amount}\\
        & \quad \SIletin{\mathit{bal}_{\mathit{dst}}}{\SIrdC{\mathit{dst}}} \\
        & \quad \color{darkgreen} // ~ \SI\mapstoCache{src}{\Cst}{v_{\mathit{src}} - \mathit{amount}} \ast \SI\mapstoCache{\mathit{dst}}{\Cst}{v_{\mathit{dst}}} ~ \land\\
        & \qquad \color{darkgreen} \mathit{bal}_{\mathit{src}} = v_{\mathit{src}} \land v_{\mathit{src}} \ge \mathit{amount} \land bal_{\mathit{dst}} = v_{\mathit{dst}}\\
        & \quad \SIwrC{\mathit{dst}}{(\mathit{bal}_{\mathit{dst}} + \mathit{amount})} \\
        & \quad \color{darkgreen} // ~ \SI\mapstoCache{\mathit{src}}{\Cst}{v_\mathit{\mathit{src}} - \mathit{amount}} \ast \SI\mapstoCache{\mathit{dst}}{\Cst}{v_\mathit{\mathit{dst}} + \mathit{\mathit{amount}}}\\
        & \qquad \color{darkgreen} \land ~ \mathit{bal}_{\mathit{src}} = v_{\mathit{src}} \land v_{\mathit{src}} \ge \mathit{amount} \land \mathit{bal}_{\mathit{dst}} = v_{\mathit{dst}}\\
        & \} \\
        & \SIcommit \\
        & \color{darkgreen} // ~ \mathit{If} ~ \mathit{commit} = \mathit{true} ~ \mathit{and} ~ v_{\mathit{src}} \ge \mathit{amount}:\\
        & \quad \color{darkgreen} \fbox{\begin{tabular}{@{}c@{}}
          $\SI\mapstoMem{\mathit{src}}{\Hist_{\mathit{src}} \listapp [v_{\mathit{src}}] \listapp [v_{\mathit{src}} - \mathit{amount}]} ~ \ast$ \\
          \hspace{-0.2cm} $\SI\mapstoMem{\mathit{dst}}{\Hist_{\mathit{dst}} \listapp [v_{\mathit{dst}}] \listapp [v_{\mathit{dst}} + \mathit{amount}]}$
          \end{tabular}} \\
        & \quad \color{darkgreen} \mathit{Otherwise}:\\
        & \quad \color{darkgreen} \fbox{$\SI\mapstoMem{\mathit{src}}{\Hist'_{\mathit{src}} \listapp [v'_{\mathit{src}}]} \ast \SI\mapstoMem{\mathit{dst}}{\Hist'_{\mathit{dst}} \listapp [v'_{\mathit{dst}}]}$}
      \end{aligned}
      \end{align*}
  \caption{Bank transfer example. \protect \footnotemark}
  \label{fig:bank-transfer-example}
\end{minipage}
\vspace{-0.2cm}
\end{wrapfigure}

\noindent of a larger invariant with multiple accounts, is simply: 
$\color{darkgreen} \exists \Hist_{\mathit{src}} ~ \Hist_{\mathit{dst}} ~ v_{\mathit{src}} ~ v_{\mathit{dst}}, ~
\SI\mapstoMem{\mathit{src}}{\Hist_{\mathit{src}} \listapp [v_{\mathit{src}}]} \ast \SI\mapstoMem{\mathit{dst}}{\Hist_{\mathit{dst}} \listapp [v_{\mathit{dst}}]}$
(it has been boxed and the existential quantification has been omitted in \Cref{fig:bank-transfer-example}).
It states that there are two accounts, named $src$ and $dst$, which points to histories 
with the latest 
updates being $v_{src}$ and $v_{dst}$ respectively 
(these are the current amounts in the accounts).
Using the invariant and \ruleref{si-start-spec}, a snapshot is 
created in \Cref{fig:bank-transfer-example} for the duration of the transaction in which the amount in the accounts 
are $v_{src}$ and $v_{dst}$. 
The if-statement is used to check whether the 
balance in the source account is 
sufficient to withdraw the amount that will be transferred (we want to avoid a negative balance). 
In case the balance is sufficiently large, the 
amount is withdrawn from the source account and inserted into the destination account 
using the \ruleref{si-write-spec} rule. 
Hence, the starting
snapshot (the local points-to resources) is changed to a snapshot 
in which the accounts contain the updated values $v_{\mathit{src}} - \mathit{amount}$ and $v_{\mathit{dst}} + \mathit{amount}$, respectively.
At commit time, it is the updates in this new snapshot that we want to push to the key-value store 
state, represented by the global points-to resources in the invariant. 
If the commit is successful, we get to do the updates atomically, i.e., write the 
updates to both accounts.
In case the commit is unsuccessful, it is because there is a conflict with another 
transaction updating the same accounts, cf.\ the snapshot isolation commit check 
captured by the $\CanCommitPredicateSymb$ predicate in \ruleref{si-commit-spec}, 
resulting in the account balances being described by fresh values. 

\footnotetext{For the sake of presentation, this example hides the use of optionals cf. the specification.}

In \citet{vMVCC}, a similar bank transfer example is proven in separation logic using 
the stronger isolation level serializability. 
As we are able to prove the bank transfer using snapshot isolation,
this example demonstrates why applications do not always have to opt for the strongest isolation levels; 
there can be a weaker isolation level, which provides sufficient consistency guarantees. 
Weaker isolation levels always come  with the benefit that they 
enhance the throughput of a system due to the increased concurrency --- something which is of 
importance in a banking system with many customers.

We remark that only the snapshot isolation specification is strong enough to prove 
the examples in this section; they can not be proven using the read uncommitted or the read committed specifications.

\section{Snapshot Isolation Implies Read Committed}
\label{sec:6}

We can now formally prove that snapshot isolation implies read committed:
\begin{theorem}[Snapshot isolation implies read committed]
  \label{si_rc_theorem}
  The specification for snapshot isolation \eqref{eq:si-setup_specs} implies the specification for read committed \eqref{eq:rc-setup_specs}: 
  \[\color{darkgreen}\mathsf{SISpec} \color{black}\rightarrow \color{darkred}\mathsf{RCSpec}.\] 
\end{theorem}
\vspace{-0.3cm}
The theorem is proven using the same structure as described in \Cref{sec:4}, and the proof amounts to approximately 
800 lines of Rocq proof code. We emphasize that while the proof follows the obvious path,
coming up with an encoding of the read committed resources using the snapshot isolation resources 
that makes the proof work is non-trivial. 
We will refer to the Rocq formalization for the details of the proof.


\section{Verifying that a Multi-Version Concurrency Control Key-Value Store Implements Snapshot Isolation}
\label{sec:7}

In this section we show that our implementation of the 
original multi-version concurrency control algorithm for snapshot isolation 
\citep{orig-SI}, as a single-node key-value store in a distributed system, 
satisfies the specification for snapshot isolation.
The section is divided into three main parts each constituting a subsection:
\begin{enumerate}
  \item \textit{Implementation}: We give an overview of the original multi-version concurrency control algorithm for snapshot isolation 
    from \citet{orig-SI} in the context of a distributed system.
  \item \textit{Client Proxy Proof}: We outline the client state and how the proof of the 
     client proxy interacts with the server-side using specifications for remote procedure call handlers.
     We include an excerpt of our ghost theory, which captures relationships among the separation logic predicates we have defined and used. 
  \item \textit{Server-Side Proof}: We go into more detail with proving remote procedure call 
    handler specifications. We include a description of the server side resources and our model 
    capturing key properties of snapshot isolation.
\end{enumerate}
We remark that we do not cover the full details of the proof, and use simplifications, as the whole proof is too large to be included 
in the paper. For the full details, see the Rocq formalization.
Note further that a formal consequence of Theorems \ref{rc_ru_theorem} and \ref{si_rc_theorem} is that the multi-version concurrency control implementation of snapshot isolation also satisfies the specifications for read committed and read uncommitted.
\vspace{-0.2cm}
\subsection{Implementation} 
\label{sec:7.1}
The data structure of the key-value store implementation is a map from keys to lists of value-timestamp pairs.
The value-timestamp pairs correspond to the values written in the past with associated commit times, with the last value being the most recent.
The server state, moreover, consists of an integer reference, for generating new timestamps to transactions, and a lock for guarding concurrent access 
to the key-value map and the timestamp integer.
The server implements three remote procedure call handlers, in the form of start, read and commit, 
which together with client proxy code makes up the start, read and commit operations that clients can use.
The write operation has no remote procedure call as it is handled locally at the client proxy.
The client proxy state can be in one of two modes: The state is empty, because no transaction is active, 
or the state contains the cached writes and the start timestamp of the active transaction.
Note that at most one transaction can be active at a time per client proxy.
For a client to have multiple active transactions, a client connection must be initialized using \ruleref{si-init-client-spec}
for each transaction. In fact, the client proxy state we talk about in this section is actually per connection.
Upon using the start operation, the client proxy asks the server to provide it with a start timestamp.
The server serves this by incrementing its timestamp counter and returning a fresh value.
Having started a transaction, the read and write operations can now be invoked.
In its implementation of the write operation, all the client proxy does is to record the update in its cache.
The read operation is more complicated.
First, it is checked whether the key, on which the read operation is invoked, has an update in the client cache.
If there is an update, the update is returned, otherwise, the read handler of the server is asked to provide a value for the key.
Together with the key, the read handler is called with the start timestamp of the transaction, as the reading must be done according to the start snapshot.
The start snapshot is represented by the start timestamp.
The server uses the key and starting timestamp to retrieve the correct value by going back into the list of value-timestamp pairs.
The value with the largest timestamp strictly less than the starting timestamp is returned.
Upon commit time, when the commit operation is invoked on the client proxy, the client proxy invokes the commit handler on the server by providing the start timestamp and all the cached writes of the transaction.
The commit handler checks that for all the keys in the cache, no updates have been made after the starting timestamp (this is the snapshot isolation commit check, cf. \Cref{fig:can_commit}).
If the check goes well, all the cached updates are appended to the lists of value-timestamp pairs.
The timestamp, in a newly added pair, will be a fresh timestamp from the server state higher than all previous timestamps.

In the following two sections we will provide more detailed explanations of the implementation
when needed to describe proof details.

\subsection{Client Proxy Proof}
\label{sec:7.2}

The client proxy proof serves as a middle layer between the top level 
specifications of \Cref{fig:specs_si} and the server side and makes use of remote procedure call handler 
specifications. When a client invokes the start, read, write or commit operation, 
it is the client proxy that gets called. Therefore, it is the client proxy who receives
the resources in the precondition of the specifications, and it is also the client proxy who
is ultimately responsible for returning the resources in the postcondition. 

\paragraph{Client-Side Resources} Upon gaining the resources in the precondition of the specifications in \Cref{fig:specs_si}, the client proxy takes a lock guarding its state and resources.
The lock is used to prevent undefined behavior if the client proxy is used concurrently by multiple threads of the client.
The lock is implemented as a spin lock, similarly to what is presented in \citet[Section 8.6]{gentleiris}.
The essence of the specifications for the acquire and release operations of the lock is that one gets to have ownership of some resources protected by the lock.
Acquire takes ownership of the resources, and release gives back the ownership to the lock for others to take.
The structure of the resources protected by the client proxy lock is displayed \eqref{eq:proxy-lock} ("\dots" is notation for resources which we have omitted for presentation purposes).
The resources consist of a pointer $\loc$, which points to the client state.
The client state is either active, pointing to a starting snapshot $\Timestamp$ and a client cache $\CacheClient$, or the state is in the inactive mode and points to nothing.
It is by using the client proxy resources in the lock, together with the resources from the preconditions of the top level specifications that the client proxy gathers the necessary resources for communicating with the server.
\begin{equation}
  \small
    Lock_{\mathit{client}}(\loc) \eqdef \exists ~ \StateClient. ~ \loc \gmapsto \StateClient \ast
      \big((\StateClient = \SOME ~ (\Timestamp, \CacheClient)  \ast \dots) \lor (\StateClient = \NONE \ast \dots)\big)
  \label{eq:proxy-lock}
\end{equation}
\paragraph{Network Communication} The network model of Aneris is an unreliable network based on UDP entailing that messages can be dropped, duplicated and reordered.
We have abstracted away from dealing with the low-level details of an unreliable network by utilizing the remote procedure call library of \citet{rel-comm}, which implements reliable communication channels on top of the unreliable network.
Hence we can think of the client-server communication as the client invoking handlers provided by the server.
The interaction between the client and the server in the proof will therefore be happening through the specification of the handlers.
Specifications corresponding to \ruleref{si-init-client-spec} and \ruleref{si-init-kvs-spec} come directly, for free, from the reliable communication library, and the write operation does not communicate with the server.
Therefore, we have provided handlers for the start, read and commit operations.
In \Cref{fig:read-handler}, we present the specification for the read handler.
The read handler is the simplest of the three handlers, as the start and commit handler has preconditions consisting of multiple non-trivial view shifts
(formally $\prop \wand \pvs[][] Q$ is view shift from $\prop$ to $Q$). 
View shifts in the precondition of handlers forces the client proxy to prove that given resources $\prop$, the server can update $\prop$ to 
the resources $Q$.
To explain the read handler specification, we first introduce our ghost theory.
\begin{figure}[ht]
  \vspace{-0.2cm}
  \small
  \centering
  \begin{mathpar}
    \axiomH{Ht-read-handler}
    {
      \anhoareV
      {
        \begin{array}[t]{@{}l@{}}
          \TimeFrag{\Timestamp} \ast 
          \SnapshotsFrag{\Timestamp}{\Memory_t} \ast 
          \MapLookup{\Memory_t}{\Key} = \SOME ~ \HistBig
        \end{array}
      }
      {
        \readHandlerC{\Key}{\Timestamp}
      }
      {
        \begin{array}[t]{@{}l@{}}
          \Ret \ValueOption. \big(\ValueOption = \NONE \ast \HistBig = \HistEmpty \big) 
            \lor \big(\exists ~ \val, \Timestamp'. ~ \ValueOption = \SOME ~ \val \ast 
              \HistVal{\HistBig} = (\val, \Timestamp')\big)
        \end{array}
      }
      {
        \srvsa
      }
      {}
    }
  \end{mathpar}
  \caption{Read handler specification.}
  \label{fig:read-handler}
  \vspace{-0.2cm}
\end{figure}
\paragraph{Ghost Theory} 
Similar to the proof sketch in \Cref{sec:4}, we will be working with a ghost theory 
of separation logic predicates together with rules that relate the different predicates.
Our ghost theory consists, among others, of the following resources: $\TimeGlobal{\Timestamp}$, $\TimeLocal{\Timestamp}$, and $\TimeFrag{\Timestamp}$, all parameterized by a 
timestamp $\Timestamp$; $\SnapshotsFrag{\Timestamp}{\Memory_t}$ and $\SnapshotsAuth{\Snapshots}$ parameterized by a 
timestamp $\Timestamp$ and a map from keys to histories $\Memory_t$ and a map $\Snapshots$ from
timestamps to maps from keys to histories; 
$\MemGlobal{\Memory}$ and $\MemLocal{\Memory}$, parameterized by a map $\Memory$ from keys to histories; and $\mapstoMem{\Key}{\HistBig}$ parameterized by a key $\Key$ and a history $\HistBig$.
Note that in this section histories are different from the terminology used in \Cref{sec:5}.
In this section, histories are lists of value-timestamp pairs as opposed to lists of values (in the client-facing specs in \Cref{sec:5} there is no need to use timestamps).
The resource $\mapstoMem{\Key}{\HistBig}$ is also different from the key-value store points-to resources in \Cref{sec:5}, but it is ultimately $\mapstoMem{\Key}{\HistBig}$ that makes up the key-value store points-to resources we have seen in \Cref{sec:5}.
Therefore, in this section we will think of $\mapstoMem{\Key}{\HistBig}$ as the key-value store points-to resources used in the \ruleref{si-start-spec} and \ruleref{si-commit-spec} rules.
An excerpt of the laws of our ghost theory are shown in \Cref{fig:ghost-theory}.
Going forward, we will go into details with our ghost theory laws as needed for explaining proof details.
\begin{figure}[h]
  \vspace{-0.2cm}
  \footnotesize
  \begin{align*}
    \TimeGlobal{\Timestamp} \ast \TimeLocal{\Timestamp} \ast \Timestamp < \Timestamp' & \proves
      \TimeGlobal{\Timestamp'} \ast \TimeLocal{\Timestamp'} \ast \TimeFrag{\Timestamp'}
    & \text{(Law 1)} \\
    \TimeGlobal{\Timestamp} \ast \TimeLocal{\Timestamp'} & \proves \TimeGlobal{\Timestamp} 
      \ast \TimeLocal{\Timestamp'} \ast \Timestamp = \Timestamp'
    & \text{(Law 2)} \\
    \MemGlobal{\Memory} \ast \MemLocal{\Memory'} & \proves \MemGlobal{\Memory} \ast 
      \MemLocal{\Memory'} \ast \Memory = \Memory'
    & \text{(Law 3)} \\
    \mapstoMem{\Key}{\HistBig'} \ast \MemGlobal{\Memory} \asts \MemLocal{\Memory} & \proves 
      \mapstoMem{\Key}{\HistBig} \ast \MemGlobal{\MapLookup{\Memory}{\mapstoMem{\Key}{\HistBig}}} \asts 
      \MemLocal{\MapLookup{\Memory}{\mapstoMem{\Key}{\HistBig}}}
    & \text{(Law 4)} \\
    \SnapshotsFrag{\Timestamp}{\Memory_t} \ast \SnapshotsAuth{\Snapshots} & \proves 
    \SnapshotsAuth{\Snapshots} \ast \MapLookup{\Snapshots}{\Timestamp} = \Memory_t
    & \text{(Law 5)} \\
    \SnapshotsAuth{\Snapshots} \ast \Timestamp \not\in \dom(\Snapshots) & \proves 
    \SnapshotsAuth{\MapLookup{\Snapshots}{\Timestamp \mapsto \Memory}} \ast \SnapshotsFrag{\Timestamp}{\Memory}
    & \text{(Law 6)}
  \end{align*}
  \vspace{-0.4cm}
  \caption{Excerpt of ghost theory laws.}
  \label{fig:ghost-theory}
  \vspace{-0.2cm}
\end{figure}
Let us now return our attention back to the read handler. The precondition
includes some client state: The transaction has a starting timestamp $\Timestamp$
with associated starting snapshot $\Memory_t$. Furthermore, the client knows 
that the key $\Key$ has the history $\HistBig$ in this snapshot.
The postcondition expresses that the read handler must return a value in accordance with the 
history $\HistBig$.

Next, we turn our attention to the server side to, among other 
things, see how the read-handler specification is proven.

\subsection{Server-Side Proof}
\label{sec:7.3}

The server-side proof is exactly the proof that the three remote procedure call handlers
(start, read and commit) implement their specifications. 
As soon as one of the handlers gets invoked, 
they will have to acquire the server lock before proceeding.

\paragraph{Server-Side Resources} The server lock resources are displayed \eqref{eq:server-lock}.
Similarly to the client lock, the server lock has memory pointers to local state: $\loc_{\kvs}$ points the key-value store itself ($\kvs$ maps keys to lists of value-timestamp pairs) while $\loc_{\Timestamp}$ points to the integer/timestamp variable used to generate start and commit timestamps.
Furthermore, the lock resources includes the timestamp resource $\TimeLocal{\Timestamp}$ and the memory resource $\MemLocal{\Memory}$.
Here $\isMap{\kvs}{\Memory}$ asserts that the ghost map $\Memory$, defined in the logic, corresponds to $\kvs$, which represent the key-value store in the program.
Lastly, we assert that our "model" holds, i.e.,
$\Model{\Memory}{\Snapshots}{\Timestamp}$ for the snapshots $\Snapshots$ ($\Snapshots$ is a map from timestamps to maps from keys to histories).
\begin{equation}
  \small
  \begin{array}{c}
      Lock_{server}(\loc_{\kvs}, \loc_{\Timestamp}) \eqdef 
      \exists ~ \kvs, \Timestamp, \map, \Memory, \Snapshots.
      \loc_{\kvs} \gmapsto \kvs \ast
      \loc_{\Timestamp} \gmapsto \Timestamp \ast 
      \TimeLocal{\Timestamp} \ast
      \MemLocal{\Memory} \asts \\
      \isMap{\kvs}{\Memory} \ast
      \Model{\Memory}{\Snapshots}{\Timestamp}
  \end{array}
  \label{eq:server-lock}
\end{equation}

\paragraph{Mathematical Model} As part of our verification effort, we have defined a mathematical model to reflect key properties of snapshot isolation in the state of the system.
The state of the system is here captured by the logical representation of the key-value store $M$, all the snapshots captured in $\Snapshots$ and the latest timestamp $\Timestamp$.
The model does not own any separation logic resources, it consists purely of logical facts (i.e., the $\Model{\Memory}{\Snapshots}{\Timestamp}$ predicate is a persistent predicate).
An important property of our model is that it defines a notion of a \emph{cut} for the state,
see the definition \eqref{eq:cut}.
\begin{equation}
  \small
    \CutC{\Timestamp}{\HistBig^t_k}{\HistBig_k} \eqdef \exists \HistBig'_k, ~ \HistBig_k = \HistBig^t_k \listapp \HistBig'_k 
      \land \left(\forall \left(\ValMath, \Timestamp' \right) \in \HistBig^t_k, ~ \Timestamp' < \Timestamp \right) 
      \land \left(\forall \left(\ValMath, \Timestamp' \right) \in \HistBig'_k, ~ \Timestamp < \Timestamp' \right)
  \label{eq:cut}
\end{equation}
A cut captures that a history $\HistBig_k$ can be split in two parts according to a timestamp $\Timestamp$, such that all the values in the first part ($\HistBig^t_k$) have timestamps strictly less than $\Timestamp$, and all the values in the latter part ($\HistBig'_k$) have timestamps strictly greater than $\Timestamp$.
For our model, it holds that if the timestamp $\Timestamp$ is associated with a snapshot of the key-value store $\Memory_t$, and $\HistBig^t_k$ is the history in the snapshot for the key $\Key$, while $\HistBig_k$ is the current history for the key, then $\Timestamp$ will cut $\HistBig_k$ and the first part of the cut be $\HistBig^t_k$.
We remark that only start timestamps have cuts, 
as all the timestamps in histories are commit timestamps because values are written at commit time
(otherwise the strict inequalities in a cut would not make sense). Likewise, we will 
only associate starting timestamps with a snapshot of the key-value store in our proof development 
(the domain of $\Snapshots$ only has starting timestamps).
\begin{equation}
  \small
    \Model{\Memory}{\Snapshots}{\Timestamp} \proves
      \big(\MapLookup{\Snapshots}{\Timestamp'} = \Memory_{t'} \land \MapLookup{\Memory_{t'}}{\Key} = \HistBig^{t'}_k
      \land \MapLookup{\Memory}{\Key} = \HistBig_k \big) \Rightarrow \CutC{\Timestamp'}{\HistBig^{t'}_k}{\HistBig_k}
  \label{eq:model-property}
\end{equation}
This cut property of the model is formally defined \eqref{eq:model-property}, and it 
is in fact this property the proof of the read handler relies on.

\paragraph{Read Handler} The code for the read handler in \Cref{fig:server-handlers}
goes through the list of value-timestamp pairs, starting with the most recent
for the specified key $\Key$, until it finds a value that has a timestamp
strictly less than the argument timestamp $\Timestamp$. 
\begin{figure}[htb!]
  \centering
  \begin{minipage}[t]{.46\textwidth}
\begin{AnerisPL}
  let start_handler time =
    let timeNext = !time + 1 in
    time := timeNext;
    timeNext

  let rec get_last opList sTime =
    match opList with
    | None -> None
    | Some list ->
      let ((v, time), tail) = list in
      if time = sTime then assert false
      else if time < sTime then Some v
      else get_last tail
  
  let kvs_get_last timeKeyPair kvs =
    let (k, sTime) = timeKeyPair in
    get_last (kvs_get k kvs) sTime

  let read_handler kvs timeKeyPair =
    kvs_get_last timeKeyPair !kvs
\end{AnerisPL}
  \end{minipage}%
  \begin{minipage}[t]{.56\textwidth}
\begin{AnerisPL}
  let commit_handler kvs cacheData time =
    let cTime = !time + 1 in
    let (sTime, cache) = cacheData in
    if map_is_empty cache then true
    else
      let b = map_forall
          (fun k _ ->
            let opList = (map_lookup k !kvs) in
            let list = 
              if opList = None then list_nil 
              else unSOME opList 
            in
            check_key k sTime list) cache 
      in
      if b then
        time := cTime;
        kvs := update_kvs !kvs cache cTime;
        true
      else false
\end{AnerisPL}
  \end{minipage}
  \caption{Excerpt of server-side handling code.}
  \label{fig:server-handlers}
\end{figure}
In accordance with the read handler specification in \Cref{fig:read-handler}, 
we must show that this value is indeed the last value of $\HistBig$ if
$\HistBig$ is non-empty. Specifically, at this point in the proof we know that 
there exists some $\HistBig_1$ and $\HistBig_2$ s.t. $\HistBig_k$ (the current history of 
key $\Key$) is composed by $\HistBig_1$ and $\HistBig_2$
in the following way: $\HistBig_k = \HistBig_1 \listapp \HistBig_2$. Furthermore, 
we can conclude that
$\forall ~ \val, \Timestamp'. ~ (\val, \Timestamp') \in \HistBig_1 \Rightarrow 
~ \Timestamp' < \Timestamp$ and 
$\forall ~ \val, \Timestamp'. ~ (\val, \Timestamp') \in \HistBig_2 \Rightarrow 
~ \Timestamp < \Timestamp'$.
Notice how this entails  that $\Timestamp$ cuts $\HistBig_k$ with 
regard to $\HistBig_1$, that is, $\CutC{\Timestamp}{\HistBig_1}{\HistBig_k}$ holds.
As it is the value of $\HistBig_1$ which will be returned,  
the proof of the read handler specification boils down to showing that 
$\HistBig = \HistBig_1$. To reach our goal, we must use a resource from the global 
invariant, shown in \eqref{eq:global-invariant}.
\begin{equation}
  \small
  \begin{array}{c}
    \GlobalInv \eqdef \exists \Timestamp_G, \Memory_G, \Snapshots_G.
      \TimeGlobal{\Timestamp_G} \ast \MemGlobal{\Memory_G} 
      \ast \SnapshotsAuth{\Snapshots_G} \asts \\
      \Model{\Memory_G}{\Snapshots_G}{\Timestamp_G}
  \end{array}
  \label{eq:global-invariant}
\end{equation}
We have defined the global invariant to contain the resources
$\TimeGlobal{\Timestamp_G}$, $\MemGlobal{\Memory_G}$, 
$\SnapshotsAuth{\Snapshots_G}$ and the logical facts of the model 
$\Model{\Memory_G}{\Snapshots_G}{\Timestamp_G}$.
Having $\SnapshotsFrag{\Timestamp}{\Memory_t}$ in the 
read handler precondition, we can use $\SnapshotsAuth{\Snapshots_G}$
from the global invariant together with Law 5 of
our ghost theory (\Cref{fig:ghost-theory}) to conclude 
$\MapLookup{\Snapshots_G}{\Timestamp} = \Memory_t$. As we know 
$\MapLookup{\Memory}{\Key} = \HistBig_k$ ($\HistBig_k$ is the current history of $\Key$), we can use  
$\MemGlobal{\Memory_G}$ from the global invariant 
and $\MemLocal{\Memory}$ from the server lock together with Law 3 of
our ghost theory to obtain the equality $\MapLookup{\Memory_G}{\Key} = \HistBig_k$.
This gives us the logical facts needed to conclude
that $\Timestamp$ cuts $\HistBig_k$ with regard to $\HistBig$
\big($\CutC{\Timestamp}{\HistBig}{\HistBig_k}$\big) using the cut property of the model \eqref{eq:model-property}.
Thus, we have established two cuts $\CutC{\Timestamp}{\HistBig_1}{\HistBig_k}$ 
and $\CutC{\Timestamp}{\HistBig}{\HistBig_k}$.
But, if a timestamp cuts the same history with regard to two different histories, 
then the two histories must be equal, giving us $\HistBig = \HistBig_1$, as needed.

\paragraph{Start Handler} The job of the start handler in \Cref{fig:server-handlers} is 
to hand out a fresh timestamp for starting transactions. 
This is done using the timestamp reference of the server state, 
which is safe as the server lock is held upon using the start handler.
Using $\TimeGlobal{\Timestamp_G}$ from the global invariant and 
$\TimeLocal{\Timestamp}$ from the server lock, 
together with Laws 1 and 2 of our ghost theory, the timestamp 
is updated to match the server state.
Moreover, $\SnapshotsAuth{\Snapshots_G}$ in the global invariant is updated using 
Law 6 of the ghost theory, with $\SnapshotsFrag{\Timestamp}{\Snapshots_G}$ being given to the client proxy, 
such that the newly generated start timestamp points to the correct snapshot 
and $\SnapshotsFrag{\Timestamp}{\Snapshots_G}$ can be used by the client with \ruleref{Ht-read-handler} later. To satisfy the model 
constraints in the global invariant and the server lock upon returning, we must update our model. 
\Cref{fig:model-updates} defines two valid updates to the model, in this case we will be using property 1.
\begin{figure}[h]
  \vspace{-0.2cm}
  \small
  \begin{align*}
    & \Model{\Memory}{\Snapshots}{\Timestamp} \ast \Timestamp < \Timestamp' \proves
      \Model{\Memory}{\MapLookup{\Snapshots}{\Timestamp' \mapsto \Memory}}{\Timestamp'}
    & \text{(Property 1)} \\
    & \Model{\Memory}{\Snapshots}{\Timestamp} \ast \Timestamp < \Timestamp' \proves
      \Model{\UpdateKVS{\Memory}{\CacheMath}{\Timestamp'}}{\Snapshots}{\Timestamp'}
    & \text{(Property 2)}
  \end{align*}
  \vspace{-0.4cm}
  \caption{Model update properties.}
  \label{fig:model-updates}
  \vspace{-0.2cm}
\end{figure}

\paragraph{Commit Handler} The code of the commit handler \Cref{fig:server-handlers} essentially consists of two parts.
First, it is checked whether the key-value store can be updated in a sound way according to the snapshot isolation commit-check.
The call to $\mathit{check}\_\mathit{key}$ is the snapshot isolation commit-check for a single key.
The $\mathit{check}\_\mathit{key}$ function takes the starting timestamp of the transaction as an argument and tests whether there is a more recent write from another transaction in the history of the argument key, after the start timestamp.
It is this part of the code which generates the $\CanCommitSymb$ status in the \ruleref{si-commit-spec} rule.
Given that the commit-check went well, the second part of the code is responsible for updating the key-value store with the writes of the transaction.
Now, the server state must be updated in a way that is sound with regard to our model.
Luckily, model-update property 2 in \Cref{fig:model-updates} imitates the changes of the $\mathit{update}\_\mathit{kvs}$ operation used to update the key-value store. 
When the $\mathit{update}\_\mathit{kvs}$ operation updates $\kvs$ using the memory pointer in the server lock \eqref{eq:server-lock}, the resource $\MemLocal{\Memory}$ together with $\MemGlobal{\Memory_G}$ from the global invariant and the key-value store points-to resources from the precondition of the \ruleref{si-commit-spec} rule, can be updated to match the updated key-value store.
This is done using Laws 3 and 4 of our ghost theory.
Moreover, the timestamp resources in the global invariant and the server lock are updated using the same approach as in the start handler.
As all resources are updated in a way that satisfies our model, the global invariant can be closed, and the server lock is re-established and can be released upon returning.

This completes our overview of the implementation and its proof; for more details, see the accompanying Rocq formalization. 

\section{Related Work}
\label{sec:8}

\paragraph{Transactions in Separation Logic}

While this paper presents the first separation logic specifications for weak isolation levels, 
there has been earlier work on separation logic specification and verification of transactions 
achieving strict serializability \citep{orig-serializability}.
Strict serializability is serializability (transactions appear to execute one at a time) 
but with the added requirement that the ordering of committed transactions respect real time order.
GoTxn \citep{DaisyNFS} is a verified transaction library, which uses two-phase commit 
locking for concurrency control and it is verified in the Perennial separation logic 
\citep{Perennial} (which, like Aneris, is also defined on top of Iris).
GoTxn stores data on disk in a non-distributed setting and is implemented on top of GoJournal \citep{DBLP:conf/osdi/ChajedTT0KZ21} 
which provides crash safety.
Likewise, the vMVCC \citep{vMVCC} transaction library uses multi-version concurrency control to implement 
a more advanced form of concurrency control than GoTxn's approach.
\citet{vMVCC} implement an in-memory key-value store in a non-distributed setting, 
and specify and verify it in 
Rocq using the Perennial separation logic, although \citet{vMVCC} do not use Perennial's support for reasoning about durable 
storage and crash safety. 

Naturally, specifications for (strict) serializability and weak isolation levels 
are vastly different. To compare with the strongest isolation level we have specified, i.e., snapshot isolation,
we note that a specification for snapshot isolation must view the state of the database twice, 
once to see the starting snapshot on which reads are made, and once to see the commit 
snapshot on which updates are committed (technically, this is reflected in our \ruleref{si-start-spec} 
and \ruleref{si-commit-spec}, where both of the preconditions include global points-to resources).
In contrast, a specification for serializability must only view the state of the database once,
as a transaction reads and performs updates on the same snapshot (informally, serializability 
can be viewed as a special case of snapshot isolation where the start snapshot and the commit 
snapshot are the same). This difference becomes even more apparent if one tries to specify a higher-order $\SIrun$ method,
which wraps start and commit operations around a body of transaction code. Indeed, 
the specification for $\SIrun$ based on serializability in 
\citet{vMVCC} is more concise than the run specification for snapshot isolation we have proven in \Cref{sec:app_util}, due 
to the relative simplicity of serializability (In \Cref{sec:app_util} we also include
a discussion of why a run specification for 
read committed and read uncommitted would be impractical.)
Contrary to what might seem intuitive, and what
one can see in more theoretically inclined database papers, serializability does
not always imply snapshot isolation. It depends on which notion of implication
is being used. As we explain in \Cref{sec:app_impl}, the implication between the separation logic specification
of serializability and snapshot isolation is not provable.

In contrast to our specification of snapshot isolation, neither vMVCC nor GoTxn provides information 
about why a commit has not succeeded, i.e., a resource in the style of the one expressed using the $\CanCommitPredicateSymb$ 
predicate in the postcondition of \ruleref{si-commit-spec}. 
This means that assertions about commit events can not be proven using the vMVCC or GoTxn specification.

\paragraph{Closing the Gap Between High Level Specifications and Implementations.} 
Other work has also aimed at closing the gap 
between the seemingly high-level properties of isolation levels and specifications 
for runnable implementations. \citet{transactions-gotsman} gives operational specifications, 
closer to specifications of actual implementations, for a number of isolation levels, and prove 
that these are equivalent to specifications in the style of weak memory models.
\citet{Crooks17} has developed a state-based model for reasoning about transactional consistency 
which has been embedded in \TLA \citep{Soethout2021}. \TLA \citep{lamport1992hybrid} is a specification
language and model checker. A version of its specification language, PlusCal, resembles 
mainstream programming languages but still has to undergo manual translation to executable code. 
\citet{raad2018semanticssnapshotisolation} shows how a pseudo-code implementation of a local transactional library 
can implement an axiomatized model of snapshot isolation equivalent to the model 
of \citet{analysing_si}. The approach is not based on a program logic or mechanized, and can 
not reason about client programs.
\citet{10.1145/3527324} lifts linearizable data structures to serializable transactional memory objects  
using interaction trees as program representations in a mechanized Rocq framework.
There is also a line of work on checking database implementations against transaction workloads 
\citep{10.1145/3552326.3567492, 10.1145/3689742, 10.1145/3591243, 10.1145/3485546}. 
While checkers have shown to be effective at discovering bugs, they can not show the absence of bugs.

\paragraph{Verification of Distributed Systems Using Separation Logic.} 
Mechanized verification of distributed systems using higher-order separation logics have been carried out a number of times with other foci than transactions.
For example, Aneris has been used to verify distributed algorithms with high-availability and lower consistency guarantees, such as conflict-free replicated data types, \citep{abel/crdts,nieto_et_al:LIPIcs.ECOOP.2023.22} and a causally-consistent replicated key-value store \citep{DBLP:journals/pacmpl/GondelmanGNTB21}.
Grove \citep{grove} is another higher-order logic for distributed systems based on Perennial with support for reasoning about durable state and crashes in which a replicated key-value store has been verified.

\paragraph{Weak Memory} The area of weak memory is another area in which separation logic 
has been successfully used to formalize relaxed guarantees, see 
\citep{10.1145/2660193.2660243, 10.1145/2509136.2509532, kaiser_et_al:LIPIcs.ECOOP.2017.17}
with \citet{kaiser_et_al:LIPIcs.ECOOP.2017.17} being an Iris based logic.
While the guarantees of weak isolation for database transactions do 
not have a direct relationship with the guarantees in weak memory, 
both areas exhibit very relaxed behavior that makes them difficult to reason about.
In the weak memory world, one starts with an operational semantics for the memory,
whereas in the work of this paper we show how an implementation of a weak isolation level, in the form of a database, 
can satisfy our specification. 


\section{Conclusion and Future Work}
\label{sec:9}

In this paper, we have established that separation logic can be used to give modular specifications of and reason
about consistency guarantees for transactions. 
As this work has laid the theoretical foundation for reasoning about transactions
in separation logic, there are two obvious paths for extending this work.  

First, this work can be seen as a stepping stone to relating specifications for
executable code, i.e., the separation logic specifications presented here, to
well-established database theory models for reasoning about transactional
consistency, i.e., models such as the ones mentioned in \Cref{sec:1}, which
define transactional consistency using, for instance, high level operational
semantics or dependency graphs. We remark that the per-operation separation
logic specifications presented in this paper already do make a connection to
existing database theory, since we have shown how they can be used to exclude
the phenomena used to define isolation levels in the literature
\citep{orig-SI,Adya99} and 1992 ANSI SQL standard, which for many years was, and
by the SQL Standard authors still is, the de-facto standard. Making a stronger
connection to one of the mentioned models, e.g., by proving formally that our
separation logic specifications relate precisely to dependency graphs, is highly
non-trivial. To create such a formal connection, it is necessary to use an extra
layer of logical machinery, not part of the standard instantiation of Iris, such
as Trillium \citep{Trillium}, for establishing refinements of state transition
systems, or free theorems \citep{free-theorems}, for stating invariants on traces
created by emitting tags in "ghost-code".

Second, it would be interesting to see the theory applied to various transaction
systems. While the database we have implemented is indeed the first executable
database that verifiably implements a weak isolation level, and uses the
multi-version concurrency control algorithm of \cite{orig-SI}, modern databases
comes with many features; e.g., distributed transactions
\citep{Lampson1981CrashRI,saga,spanner}, where data is scattered across a number
of different databases. 
Further, it would be interesting to see the developed theory applied to 
databases that are resilient to crash failures.
This would necessitate using a distributed program logic with support for 
writing to durable storage and reasoning about crashes, such as Grove \citep{grove}.


\section*{Data Availability Statement}
The Rocq formalization and OCaml code accompanying this work 
is available on Zenodo \citep{abc}.
\section*{Acknowledgments}

This work was supported in part by a Villum Investigator grant (no. 25804), 
Center for Basic Research in Program Verification (CPV), from the VILLUM Foundation.
This work was co-funded by the European Union (ERC, CHORDS, 101096090).
Views and opinions expressed are however those of the author(s) only and
do not necessarily reflect those of the European Union or the European
Research Council. Neither the European Union nor the granting authority
can be held responsible for them.

\appendix
\section{Utility Code}
\label{sec:app_util}

In addition to the operations of the key-value store API, we have implemented 
utility operations $\SIwait$ and $\SIrun$. 
The $\SIwait$ and $\SIrun$ operations are implemented 
on top of the key-value store operations and proven correct with regard to their 
specifications (the specifications are presented in this section) using the 
specifications from \Cref{fig:specs_ru,fig:specs_rc,fig:specs_si}.
The $\SIrun$ is similar to the one presented for serializability in \citet{vMVCC}, 
but has a vastly different specification as discussed in \Cref{sec:9}. 
As we will see in the following, while the utility operations provide new functionality, 
they are good at showcasing the differences between the isolation levels.
 
\begin{figure}[htb!]
  \centering
  \begin{minipage}[t]{.37\textwidth}
  \begin{AnerisPL}
    let wait cst cond k =
      let rec aux () =
        start cst ;
        match read cst k with
        | None ->
          commit cst; aux ()
        | Some v ->
          if cond v
          then (commit cst)
          else (commit cst; aux ())
in aux ()
  \end{AnerisPL}
  \end{minipage}%
  \begin{minipage}[t]{.34\textwidth}
    \begin{AnerisPL}
    let weak_wait cst cond k =
      start cst;
      let rec aux () =
        match read cst k with
        | None -> aux ()
        | Some v ->
          if cond v
          then (commitU cst)
          else (aux ())
      in aux ()
    \end{AnerisPL}
    \end{minipage}%
  \begin{minipage}[t]{.33\textwidth}
  \begin{AnerisPL}
    let run cst handler =
      start cst;
      handler cst;
      commit cst
  \end{AnerisPL}
  \end{minipage}
  \vspace{-2em}
  \caption{Utility code.}
  \label{fig:utility-implementation}
  \vspace{-1em}
\end{figure}

\paragraph{Wait operation} 
The $\SIwait$ operation comes in two versions: a \emph{weak} version, for read uncommitted and 
read committed, and a non-weak version for snapshot isolation. 
The operation works much like a memory fence: It
takes as arguments a connection, key and condition. When the operation returns, 
the condition holds for the value pointed to by the key. 
The $\SIwait$ operation can be used to synchronize transactions across clients 
to impose a particular order. 
The $\SIwait$ and weak-$\SIwait$ are implemented as recursive functions, see \Cref{fig:utility-implementation}. 
Starting with the non-weak $\SIwait$ operation: 
Upon every call, a new transaction is started to obtain the current snapshot of the key-value store. 
The snapshot is then checked for the condition by reading from the specified 
key and seeing whether the condition holds for the value pointed to be the key. 
If the condition is true, the operation commits the transaction and terminates, 
otherwise the transaction is committed, and a recursive call is made to check 
the condition on a more recent snapshot. Notice how every call starts and commits 
a transaction. Had the $\SIwait$ operation only started one transaction and done the 
recursive calls inside this transaction, it would never get a 
new start snapshot and hence not observe changes done by other transactions, including
changes that could make the condition true. Now, this part is what distinguishes 
the non-weak $\SIwait$ from the weak $\SIwait$: The weak $\SIwait$ operation does not start a new 
transaction for every recursive call to check the state of the key. 
This is not needed for read uncommitted and read committed as 
transactions can see updates made by other transactions during their own execution. 
Do note that the key-value store does not \emph{have} to show updates to 
transactions that are already execution, 
e.g., the key-value store we have implemented in \Cref{sec:7} have realized the specifications 
for read uncommitted and read committed through the implications proofs without propagating new 
values to active transactions, 
but if you do not, then you might as well implement a stronger isolation level. 
\begin{figure}[h]
  \footnotesize
  \begin{mathpar}
    \color{darkgreen}
    \axiomH{Wait-spec}
    { 
      \anhoareV
        { 
          \begin{array}[t]{@{}l@{}}
          \SI\CanStart{\Cst} \asts \\
          \axiomH{}
          { 
            \begin{array}[t]{@{}l@{}}
              \forall \val'.
              \end{array}
            \anhoare
              { \begin{array}[t]{@{}l@{}}
                \TRUE
                \end{array}
              }
              { \begin{array}[t]{@{}l@{}}
                \SIcond ~ \val'
              \end{array}
              }
              { \begin{array}[t]{@{}l@{}}
                \Ret \Boolean. \Boolean \Rightarrow \val = \val'
              \end{array}
              }
              {}{}
          }
          \asts \\
          \always \pvs[\top][\mask]
          \Big(\exists \Hist, ~  \SI\mapstoMem{\Key}{\Hist} \ast
          \later \big(\SI\mapstoMem{\Key}{\Hist} \vsW[\mask][\top] \TRUE \big)\Big)
          \end{array}
        }
        { 
          \begin{array}[t]{@{}l@{}}
          \SIwaitC{\Cst}{\Key}{\SIcond}
          \end{array}
        }
        { 
          \begin{array}[t]{@{}l@{}}
          \Ret \TT. \SI\CanStart{\Cst} \ast \SI\Seen{\Key}{\Hist \listapp [\val]} 
          \end{array}
        }
        {}{}
    }
    \color{darkred}
    \axiomH{Weak-wait-spec}
    { 
      \anhoareV
        { 
          \begin{array}[t]{@{}l@{}}
          \RC\CanStart{\Cst} \asts \\
          \axiomH{}
          { 
            \begin{array}[t]{@{}l@{}}
              \forall \val'.
              \end{array}
            \anhoare
              { \begin{array}[t]{@{}l@{}}
                \TRUE
                \end{array}
              }
              { \begin{array}[t]{@{}l@{}}
                \SIcond ~ \val'
              \end{array}
              }
              { \begin{array}[t]{@{}l@{}}
                \Ret \Boolean. \Boolean \Rightarrow \val = \val'
              \end{array}
              }
              {}{}
          }
          \asts \\
          \always \pvs[\top][\mask]
          \Big(\exists \Vals, ~  \RC\mapstoMem{\Key}{\Vals} \ast
          \later \big(\RC\mapstoMem{\Key}{\Vals} \vsW[\mask][\top] \TRUE \big)\Big)
          \end{array}
        }
        { 
          \begin{array}[t]{@{}l@{}}
          \SIwaitC{\Cst}{\Key}{\SIcond}
          \end{array}
        }
        { 
          \begin{array}[t]{@{}l@{}}
          \Ret \TT. \RC\CanStart{\Cst} \ast \RC\Seen{\Key}{\Vals \cup \val} 
          \end{array}
        }
        {}{}
    }
  \end{mathpar}
  \vspace{-2em}
   \caption{Wait specification and weak wait specification.}
   \label{fig:wait-spec}
   \label{fig:weak-wait-spec}
   \vspace{-1em}
\end{figure}
The specification for the non-weak $\SIwait$ operation is shown in \Cref{fig:wait-spec}.
It has three resources in its precondition.
First, one must provide the ability to start a transaction in the form of $\CanStart{\Cst}$.
Second, the user must show that the provided condition is a test for equality for a specific value $\val$.
Third, the user must provide the ability to look at the state of the key-value store all the possibly many times the $\SIwait$ operation will be opening and closing transactions to check for the condition.
This is done in the form of an update modality and a view shift 
($\prop \vsW[][] Q$ is a view shift and is notation for $\prop \wand \pvs[][] Q$)
\citep{iris} under the persistence modality 
(going forward 
we will refer to both $\vsW$ and $\pvs$ as view shifts).
We recap that the update modality $\pvs[\mask_1][\mask_2] \prop$ states that given that the invariants in $\mask_1$ are active 
($\mask_1$ is a set of invariant names commonly referred to as a mask, we omitted invariant names, used as a mean to 
identify each invariant, in the main body of the paper), updates can be made to satisfy $\prop$ and establish the invariants in $\mask_2$.
If the two masks $\mask_1$ and $\mask_2$ are equal, we can choose to write the update modality as $\pvs[\mask]{}$ or simply ignore the mask $\mask$.
The persistence modality ensures that the view shift can be reused all the possibly many times the $\SIwait$ operation will be opening and closing transactions.
Specifically, the view shifts allow us, possibly by opening an invariant, to get the key-value store points-to resource for the argument key and later close any invariant using the same key-value store points-to resource.
Given the mentioned preconditions, if the $\SIwait$ operation terminates the user obtains evidence of the observation in the form of $\color{darkgreen}\SI\Seen{\Key}{\Hist \listapp [\val]}$.
One can then use the snapshot isolation version of exclusion property \eqref{eq:ru-exclusion-rule}
(see \Cref{sec:5}) to exclude future key-value store states where the value $\val$ has not been written to the key $\Key$ yet.
We remark that in our Rocq development, a more general specification that does not restrict the condition to be an equality is provided.
The specification of the weak $\SIwait$ operation is much similar to its stronger counterpart.
\Cref{fig:weak-wait-spec} displays the weak $\SIwait$ specification for read committed, we remark 
that read uncommitted has an equivalent specification (all resources should be prefixed with 
$\color{darkblue}\RU$ instead of $\color{darkred}\RC$).
The difference between the weak $\SIwait$ operation and the non-weak operation, on the specification level,
lies in the kind of observations one gains when the operation returns. 
The read committed (and read uncommitted observations) $\color{darkred}\RC\Seen{\Key}{\Vals \cup \val}$
are on sets which imposes no order compared to the histories of snapshot isolation. 
This only lets the user exclude states based on whether 
a given write have happened or not but not based on the order it happened. 

Using the non-weak $\SIwait$ specification for snapshot isolation, we have proven two examples for 
concurrently executing transactions
that make assertions based on the observations gained from using $\SIwait$.
The first example, \emph{Capturing causality} in \Cref{fig:capturing-causality-example}, uses wait to 
capture a causality relation between transactions.
The second example, \emph{Sequential writes commit} in \Cref{fig:sequential-writes-to-the-same-key-commit-example}, 
asserts that committing will be successful as, even though the two transactions write to the same key, 
there can not be a write-conflict because the second transaction is waiting for the first transaction to 
finish.
\begin{figure}[h]
  \small
  \centering
  \vspace{0.3cm}
  \begin{minipage}[b]{0.5\textwidth}
    \begin{align*}
    \left.
    \begin{aligned}
      & \SIstart\\
      & \SIwrC{x}{1}\\
      & \SIcommit\\ 
      & \\ 
      &
    \end{aligned}
    ~\middle\Vert~
    {
    \left.
    \begin{aligned}
      & \SIwaitC{x}{1}\\
      & \SIstart\\
      & \SIwrC{y}{1}\\
      & \SIcommit\\ 
      & 
    \end{aligned}
    ~\middle\Vert~
    \begin{aligned}
      & \SIwaitC{y}{1}\\
      & \SIstart\\
      & \SIletin{v_x}{\SIrdC{x}}\\
      & \SIassert{v_x = 1}\\
      & \SIcommit
    \end{aligned}
    \right.
    }
    \right.
    \end{align*}
    \begin{align*}
      Inv &\eqdef{} \color{darkgreen} \exists h_x ~ h_y,~ \SI\mapstoMem{x}{h_x} \ast \SI\mapstoMem{y}{h_y} ~ \ast \\
                    & \color{darkgreen} \big((h_x = \HistEmpty \ast h_y = \HistEmpty)
                      \lor ~ (\HistVal{h_x} = \Some 1 ) \big)
    \end{align*}
    \vspace{-0.6cm}
    \caption{Capturing causality.}
    \label{fig:capturing-causality-example}
    \vspace{-0.3cm}
  \end{minipage}\hspace*{-0.2em}%
  \begin{minipage}[b]{0.4\textwidth}
    \begin{align*}
    \left.
    \begin{aligned}
      & \SIstart\\
      & \SIwrC{x}{1}\\
      & \SIassert{\SIcommit}\\ 
      &
    \end{aligned}
    ~\middle\Vert~
    {
    \begin{aligned}
      & \SIwaitC{x}{1}\\
      & \SIstart\\
      & \SIwrC{x}{2}\\
      & \SIassert{\SIcommit}\\
    \end{aligned}
    }
    \right.
    \end{align*}
    \begin{align*}
      Inv &\eqdef{} \color{darkgreen} \exists h,~ \SI\mapstoMem{x}{h} ~ \ast \\
      & \color{darkgreen} ((h = \HistEmpty)
        \lor ~ (h = \HistNonEmpty{1} \ast \FinishToken_1)
        \lor ~ (h = \HistNonEmpty{1;2} \ast \FinishToken_1 \ast \FinishToken_2))
    \end{align*}
    \vspace{-0.6cm}
    \caption{Sequential writes commit.}
    \label{fig:sequential-writes-to-the-same-key-commit-example}
    \vspace{-0.3cm}
  \end{minipage}
\vspace{0.2cm}
\end{figure}

\paragraph{Run operation} The $\SIrun$ operation is a higher-order function, which wraps start and commit operations around a body of transaction code.
Its specification, shown in \Cref{fig:run-spec}, naturally follows from the specifications in \Cref{fig:specs_si}: To be able to start a transaction, the \ruleref{Run-spec} rule must have $\CanStart{\Cst}$ and a snapshot of the key-value store.
\begin{figure}[h]
  \small
  \centering
  \begin{mathpar}
    \color{darkgreen}
    \axiomH{Run-spec}
    { 
      \anhoareV
        { 
          \begin{array}[t]{@{}l@{}}
            \SI\CanStart{\Cst} ~ \asts \\
            \pvs[\top][\mask]
            \Big(\exists \map, ~ \Ppred{\map} \ast \Sep_{(\Key,\, \Hist) \in\, \map} \SI\mapstoMem{\Key}{\Hist} \ast
            \later \big(\Sep_{(\Key,\, \Hist) \in\, \map} \SI\mapstoMem{\Key}{\Hist} \vsW[\mask][\top] \TRUE \big)\Big)
            ~ \asts \\
            \forall \Snapshot. ~ \anhoareV
              {
                \begin{array}[t]{@{}l@{}}
                  \Sep_{(\Key,\, \Hist) \in\, \Snapshot}
                  \big(\SI\mapstoCache{\Key}{\Cst}{\HistVal{\Hist}} \asts \SI\KeyUpdStatus{\Cst}{\Key}{\FALSE}\big)
                  \ast \Ppred{\Snapshot}
                \end{array}
              }
              {
                \begin{array}[t]{@{}l@{}}
                  \Body ~ \Cst
                \end{array}
              }
              {
                \begin{array}[t]{@{}l@{}}
                  \Ret \TT. ~ \dom ~ \Snapshot =  \dom ~ \Cache \asts \\
                  \Sep_{(\Key,\, (\ValueOption,\, \Boolean)) \in\, \Cache} 
                  \big(\SI\mapstoCache{\Key}{\Cst}{\ValueOption} \ast \SI\KeyUpdStatus{\Cst}{\Key}{\Boolean} \big) \asts \\
                  \pvs[\top][\mask]
                  \Big(\exists \map, ~ Q(\map, \Snapshot, \Cache) \ast \Sep_{(\Key,\, \Hist) \in\, \map} \SI\mapstoMem{\Key}{\Hist}
                  \ast \dom ~ \map = \dom ~ \Snapshot \asts \\
                  \hspace{1.3cm} \later 
                  \big((\Sep_{\substack{(\Key,\, \Hist) \in\, \map \\ (\Key,\, p) \in\, \Cache}} \SI\mapstoMem{\Key}{\CommitHist{p}{\Hist}} \lor
                  \Sep_{(\Key,\, \Hist) \in\, \map} \SI\mapstoMem{\Key}{\Hist})
                  \vsW[\mask][\top] \TRUE \big)\Big)
                \end{array}
              }
              {}{}
          \end{array}
        }
        { 
          \begin{array}[t]{@{}l@{}}
          \SIrunC{\Cst}{\Body}
          \end{array}
        }
        { 
          \begin{array}[t]{@{}l@{}}
            \Ret \val. \SI\CanStart{\Cst} \ast Q(\map, \Snapshot, \Cache) \asts \\
            \Big(\val = \TRUE \ast \SI\CanCommit{\map}{\Snapshot}{\Cache} \ast 
            \hspace{0.2cm} \Sep_{\substack{(\Key,\, \Hist) \in\, \map \\ (\Key,\, p) \in\, \Cache}}
            \SI\Seen{\Key}{\SI\CommitHist{p}{\Hist}}\Big) \\
            \lor ~ \Big(\val = \FALSE \ast \lnot \SI\CanCommit{\map}{\Snapshot}{\Cache} ~ \asts
            \Sep_{(\Key,\, \Hist) \in\, \map} \SI\Seen{\Key}{\Hist}\Big)
          \end{array}
        }
        {}{}
    }
  \end{mathpar}
  \vspace{-2em}
   \caption{Run specification.} 
   \label{fig:run-spec}
   \vspace{-1em}
\end{figure}
The view shifts used for taking a snapshot of the key-value store follows the same idea as in the precondition of the $\SIwait$ specification.
The differences are that the persistence modality is not needed, as the view shift is only used once to start the transaction, and also that the snapshot can be described by a user-defined predicate $\prop$.
Having started the transaction, the $\SIrun$ specification will have to reason about the body of the transaction based on the snapshot described by $\prop$.
As the body of the transaction is provided by the user of the specification, the \ruleref{Run-spec} rule includes in its precondition a Hoare triple for the transaction body.
The Hoare triple for the transaction body has itself as precondition connection points-to resources and update status resources corresponding to the ones obtained from starting a transaction using the first two resources of the \ruleref{Run-spec} rule precondition.
From using the transaction body Hoare Triple, the connection points-to and updates status resources potentially gets modified.
The changes are captured in the map $\Cache$ in the postcondition of the triple.
For committing the transaction, based on the modifications in $\Cache$ and the snapshot $\Snapshot$, the \ruleref{Run-spec} rule uses the two view shifts it also obtains from using the transaction body Hoare triple.
These view shifts describe exactly how one can obtain, possible by opening invariants, a commit snapshot of the key-value store and close any invariants using one of two snapshots: The unmodified commit snapshot corresponding to the unsuccessful commit case, or the commit snapshot with the updates described by $\Cache$ corresponding to the successful commit case.
The postcondition of the \ruleref{Run-spec} rule reflects the commit status of the transaction body: The return status of the $\SIrun$ operation is attached to $\CanCommitSymb$ expressed using a commit snapshot $\map$, a start snapshot $\Snapshot$ and modifications $\Cache$.
All three maps are related by the user-provided predicate $Q$, just as in the postcondition of the transaction body.
For excluding future key-value store states, $\SeenSymb$ predicates are also returned based on the commit status of the transaction body.

While the $\SIrun$ specification for snapshot isolation is complex, and most likely it is more convenient to 
reason about transactions running under snapshot isolation using the per operation specifications of 
\Cref{fig:specs_si}, the $\SIrun$ specifications for read committed and read uncommitted would be far 
more complex. As the specifications of read and write access shared key-value store state in 
read uncommitted and read committed, a $\SIrun$ specification 
must take as argument resources in the form of view shifts 
that can be used to reason about the access to shared key-value store state for all the operations 
of the transaction body --- not only the start and commit operations as in the $\SIrun$ specification 
for snapshot isolation \Cref{fig:run-spec}.
\section{Example Portfolio}
\label{sec:app_portfolio}

In addition to the examples presented in the body of the paper, we have verified 
a number of additional examples for the strongest isolation level of snapshot isolation
displaying important properties of snapshot isolation.
All the examples have been proven in Rocq using our specifications in \Cref{fig:specs_si}
and can be found in the accompanying Rocq files.
The invariant is listed together with each example, unless an invariant was not needed.
This is contrary to the body of the paper where we sometimes omit invariants for presentation 
purposes. 
In the following, we will attach some comments to each of the examples.

The \emph{atomic transactions} example in \Cref{fig:atomic-transactions} 
xpress that transactions are atomic using an assert statement: 
Either all writes in a transaction are committed, or no writes are committed --- 
it is not possible to read one value from one transaction and another value from the other transaction.
 
The \emph{convenience of points-to} example in \Cref{fig:convenience-of-points-to-example}
shows the convenience of working with 
points-to predicates as it allows for modularity: One can make a specification for the 
function $f$ that only mentions the points-to-predicate for the key it takes as argument.

The \emph{disjoint writes} example in \Cref{fig:disjoint-writes-commit-example} asserts that transactions writing to disjoint sets of keys will commit successfully.
This holds as the snapshot isolation commit-check in \Cref{fig:can_commit} is only concerned with write-conflicts which can not be created from disjoint writes.
Furthermore, the disjoint writes example shows the modularity of the specs: No invariant is needed as the state of the key-value store is expressed using points-to predicates.
This allows us to distribute the points-to resources for keys $x$ and $y$ to the left and right transactions respectively.

The \emph{read-only transaction} example in \Cref{fig:read-transactions-commit-example} asserts that a read-only transaction will commit successfully.
As mentioned before, this holds because the snapshot isolation commit-check is only concerned with write-conflicts.

The last example in \Cref{fig:difference-example} displays the difference between snapshot isolation and serializability: Snapshot isolation is a weaker property than serializability as any execution of committed transactions allowed under serializability is also allowed under snapshot isolation.
We can think of serializability as a special case of snapshot isolation where the start snapshot and the commit snapshot is the same.
Now, this comes with the caveat that snapshot isolation guarantees that all transactions will be reading from a valid snapshot, while serializability gives no guarantees about transactions that do not successfully commit.
The implication of this is that under snapshot isolation we can make assertions before a transaction has committed.
If we refrain from considering that assertions must be made after successfully committing in serializability, we can not create an example, which is only provable under snapshot isolation as an algorithm implementing snapshot isolation can then behave exactly as one implementing serializability.
To illustrate this point, observe that the left case in the disjunction of the assertion in \Cref{fig:difference-example}, $v_x + v_y = \text{-}2$, corresponds to the only execution not allowed under serializability.
Namely, the execution in which the writing transactions have the same start snapshot.
Had this case not been included in the assertion, we would only have been able to prove the (remaining) assertion for an algorithm implementing serializability, but not for one implementing snapshot isolation.

\begin{figure}[h]
  \small
  \centering
    \begin{minipage}[b]{0.5\textwidth}
    \begin{align*}
    \left.
    \begin{aligned}
      & \SIstart\\
      & \SIwrC{x}{1}\\
      & \SIwrC{y}{1}\\
      & \SIcommit\\
      &
    \end{aligned}
    ~\middle\Vert~
    {
    \left.
    \begin{aligned}
      & \SIstart\\
      & \SIwrC{x}{2}\\
      & \SIwrC{y}{2}\\
      & \SIcommit\\
      &
    \end{aligned}
    ~\middle\Vert~
    \begin{aligned}
      & \SIstart\\
      & \SIletin{v_x}{\SIrdC{x}}\\
      & \SIletin{v_y}{\SIrdC{y}}\\
      & \SIassert{v_x = v_y}\\
      & \SIassert{\SIcommit}
    \end{aligned}
    \right.
    }
    \right.
    \end{align*}
    \begin{align*}
      Inv &\eqdef{} \color{darkgreen} \exists h,~ \SI\mapstoMem{x}{h} \ast \SI\mapstoMem{y}{h}
    \end{align*}
    \vspace{-0.6cm}
    \caption{Atomic transactions.}
    \label{fig:atomic-transactions}
    \vspace{-0.3cm}
  \end{minipage}%
    \begin{minipage}[b]{0.5\textwidth}
    \centering
    \begin{align*}
    \left.
    \begin{aligned}
      & \SIstart\\
      & \SIwrC{x}{1}\\
      & \SIcommit\\ 
      &
    \end{aligned}
    ~\middle\Vert~
    {
     \begin{aligned}
      & \SIstart\\
      & \SIletin{r}{f(x)}\\
      & \SIwrC{y}{r}\\
      & \SIcommit 
    \end{aligned}
    }
    \right.
    \end{align*}
    \begin{align*}
      Inv &\eqdef{} \color{darkgreen} \exists h,~ \SI\mapstoMem{x}{h}
    \end{align*}
    \vspace{-0.6cm}
    \caption{The convenience of points-to.}
    \label{fig:convenience-of-points-to-example}
    \vspace{-0.3cm}
  \end{minipage}
  \vspace{-1em}
\end{figure}
\begin{figure}[h]
  \small
  \centering
    \begin{minipage}[b]{0.5\textwidth}
    \begin{align*}
    \left.
    \begin{aligned}
      & \SIstart\\
      & \SIwrC{x}{1}\\
      & \SIassert{\SIcommit}
    \end{aligned}
    ~\middle\Vert~
    {
    \begin{aligned}
      & \SIstart\\
      & \SIwrC{y}{1}\\
      & \SIassert{\SIcommit}
    \end{aligned}
    }
    \right.
    \end{align*}
    \vspace{-0.4cm}
    \caption{Disjoint writes commit.}
    \label{fig:disjoint-writes-commit-example}
  \end{minipage}%
  \begin{minipage}[b]{0.5\textwidth}
    \begin{align*}
    \left.
    \begin{aligned}
      & \SIstart\\
      & \SIwrC{x}{1}\\
      & \SIcommit
    \end{aligned}
    ~\middle\Vert~
    {
    \begin{aligned}
      & \SIstart\\
      & \SIrdC{x}\\
      & \SIassert{\SIcommit}
    \end{aligned}
    }
    \right.
    \end{align*}
    \begin{align*}
      Inv &\eqdef{} \color{darkgreen} \exists h,~ \SI\mapstoMem{x}{h}
    \end{align*}
    \vspace{-0.6cm}
    \caption{Read-only transactions commit.}
    \label{fig:read-transactions-commit-example}    
  \end{minipage}%
  \vspace{-1em}  
\end{figure}
\begin{figure}[h]
  \small
  \centering
   \begin{minipage}{1\textwidth}
    \begin{align*}
    \left.
    \begin{aligned}
      & \SIstart\\
      & \SIwrC{x}{1}\\
      & \SIwrC{y}{1}\\
      & \SIwrC{z}{1}\\
      & \SIcommit\\ 
      &
    \end{aligned}
    ~\middle\Vert~
    {
    \left.
    \begin{aligned}
      & \SIwaitC{z}{1}\\
      & \SIstart\\
      & \SIletin{v_x}{\SIrdC{x}}\\
      & \SIifthen{(v_x = 1)}\\ 
      & \{\SIwrC{y}{(\text{-}1)}\}\\
      & \SIcommit 
    \end{aligned}
    ~\middle\Vert~
    {
    \left.
    \begin{aligned}
      & \SIwaitC{z}{1}\\
      & \SIstart\\
      & \SIletin{v_y}{\SIrdC{y}}\\
      & \SIifthen{(v_y = 1)}\\
      & \{\SIwrC{x}{(\text{-}1)}\}\\
      & \SIcommit  
    \end{aligned}
    ~\middle\Vert~
    \begin{aligned}
      & \SIwaitC{z}{1}\\
      & \SIstart\\
      & \SIletin{v_x}{\SIrdC{x}}\\
      & \SIletin{v_y}{\SIrdC{y}}\\
      & \SIassert{v_x + v_y = \text{-}2 \lor v_x + v_y \geq 0}\\
      & \SIcommit
    \end{aligned}
    \right.
    }
    \right.
    }
    \right.
    \end{align*}
    \begin{align*}
      & Inv \eqdef{} \color{darkgreen} \exists h_x ~ h_y ~ h_z,~ \SI\mapstoMem{x}{h_x} \ast \SI\mapstoMem{y}{h_y} \ast
                  \SI\mapstoMem{z}{h_z} ~ \ast \\
      & \hspace{-0.6em} \color{darkgreen} \quad \Big(\big(h_x = \HistEmpty \big) \lor
        \big(\exists ~ \val_x, \val_y. ~ \HistVal{h_x} = \Some \val_x \ast \HistVal{h_y} = \Some \val_y \asts
        (\val_x = 1 \lor \val_x = \text{-}1) \ast
        (\val_y = 1 \lor \val_y = \text{-}1) \big)\Big)
    \end{align*}
    \vspace{-0.6cm}
    \caption{Capturing the difference between snapshot isolation and serializability.}
    \label{fig:difference-example}
    \vspace{-1em}
  \end{minipage}
\end{figure}

\section{Serializability Does not Imply Snapshot Isolation in Separation Logic}
\label{sec:app_impl}

Consider an example of two transactions with a single write operation
both writing to the same key. All executions for these transactions are valid
under serializeability. For snapshot isolation, only some executions are: the
execution in which the two transactions share the same starting snapshot is not.
The notion of implication we use (which is the natural one when thinking of
specifications in terms of Hoare Triples), requires us to show that any
execution valid for serializability is also valid under snapshot isolation.
Hence, the implication can not be proven. In other papers, the implication is
sometimes interpreted as: if there exists an execution valid for serilizability,
then there must exist some execution (for the same set of transaction of course)
which is valid under snapshot isolation. Hence, using this definition the
implication can be proven.


\bibliography{paper}
\end{document}